\documentclass[aps,twocolumn,
superscriptaddress,
footinbib,
pra, 10pt, longbibliography, floatfix]{revtex4-2}

\usepackage{graphicx}
\usepackage{enumitem}
\usepackage{mwe}
\usepackage{mathtools}
\usepackage{bbold}
\usepackage{amsmath,amssymb,bm}
\usepackage{mathrsfs}
\usepackage{MnSymbol}
\usepackage{graphicx}
\usepackage[caption=false]{subfig}
\DeclareMathOperator*{\argmax}{arg\,max}
\usepackage{comment}
\usepackage[usenames,dvipsnames]{color}
\usepackage{natbib}
\usepackage{xcolor, colortbl}
\usepackage{physics}
\usepackage{subfig}
\usepackage{tikz}
\usetikzlibrary{quantikz2}
\usepackage{diagbox}
\usepackage{hyperref}
\usepackage[nameinlink]{cleveref}

\usepackage{natbib}

\begin{document}
\title{Limitations for adaptive quantum state tomography in the presence of detector noise}

\date{\today}

\author{Adrian Skasberg Aasen}
\email{adrian.aasen@uni-jena.de}

\affiliation{Institut für Festk\"{o}rpertheorie und -optik,  Friedrich-Schiller-Universit\"{a}t Jena, Max-Wien-Platz 1, 07743 Jena, Germany}
\affiliation{Kirchhoff-Institut f\"{u}r Physik, Universit\"{a}t Heidelberg, Im Neuenheimer Feld 227, 69120 Heidelberg, Germany}

\author{Martin G\"{a}rttner}
\email{martin.gaerttner@uni-jena.de}
\affiliation{Institut für Festk\"{o}rpertheorie und -optik,  Friedrich-Schiller-Universit\"{a}t Jena, Max-Wien-Platz 1, 07743 Jena, Germany}

\begin{abstract}
Assumption-free reconstruction of quantum states from measurements is essential for benchmarking and certifying quantum devices, but it remains difficult due to the extensive measurement statistics and experimental resources it demands. An approach to alleviating these demands is provided by adaptive measurement strategies, which can yield up to a quadratic improvement in reconstruction accuracy for pure states by dynamically optimizing measurement settings during data acquisition. A key open question is whether these asymptotic advantages remain in realistic experiments, where readout is inevitably noisy. In this work, we analyze the impact of readout noise on adaptive quantum state tomography with readout-error mitigation, focusing on the challenging regime of reconstructing pure states using mixed-state estimators. Using analytical arguments based on Fisher information optimization and extensive numerical simulations using Bayesian inference, we show that any nonzero readout noise eliminates the asymptotic quadratic scaling advantage of adaptive strategies. We numerically investigate the behavior for finite measurement statistics for single- and two-qubit systems with exact readout-error mitigation and find a gradual transition from ideal to sub-optimal scaling. We furthermore investigate scenarios where detector tomography is performed with a limited number of state copies for calibration, showing that insufficient detector characterization leads to estimator bias and limited reconstruction accuracy. Although our result imposes an upper bound on the reconstruction accuracy that can be achieved with adaptive strategies, we observe numerically a constant-factor gain in reconstruction accuracy, which becomes larger as the readout noise decreases. This indicates potential practical benefits in using adaptive measurement strategies in well-calibrated experiments.

\end{abstract}

\maketitle

\section{Introduction}
\label{sec:introduction}



Quantum state tomography is the most measurement intensive characterization and benchmarking tool for quantum systems. Consequently, extensive research efforts have been devoted to identifying optimal measurement strategies that use fewer state copies to achieve the desired reconstruction accuracy. Collective measurement, where $N$ copies of a state are jointly measured, can achieve optimal asymptotic scaling for reconstruction infidelity of $1/N$ for arbitrary quantum states \cite{Massar1995, Bagan2006B, Haah2016, pelecanos2025}. Among static (nonadaptive) copy-by-copy measurement schemes,  one near-optimal approach is to measure mutually unbiased bases \cite{Wootters1989, Mahler2013, adamson2010}. Static approaches are particularly effective for strongly mixed states, where no eigenvalues are close to zero, generally achieving an asymptotic infidelity that scales as $1/N$ \cite{Kahn2009}. 
Throughout this work, we focus exclusively on copy-by-copy measurement schemes, and therefore we will use the terms state copies and measurements interchangeably.

Pure and close-to-pure states are particularly difficult to reconstruct because they are close to the boundary of the estimation domain \cite{Mahler2013, Struchalin2018, Bogdanov2009, Haah2016}. These states are of practical concern, as most quantum experiments intend to produce pure states. Compared to strongly mixed states, pure states show a quadratically poorer asymptotic infidelity scaling for static copy-by-copy measurement schemes, going from $1/N \rightarrow 1/\sqrt{N}$ \cite{Haah2016, Kahn2009, Struchalin2018, Bogdanov2009}. Adaptive measurement strategies mend this discrepancy by introducing classical communication between measurements, which allows for higher information extraction per state copy. If projective measurements can be performed on multiqubit entangled bases, optimal reconstruction infidelity is recovered with a single adaptive step in the middle of the measurement process \cite{Struchalin2018, Mahler2013}. If restricted to more practical separable projective measurement, continuous adaptive optimizations are generally required to achieve the best asymptotic scaling for arbitrary pure states \cite{Huszr2012, Struchalin2018, Struchalin2016, Straupe2016, Sugiyama2012}.  The current state-of-the-art adaptive quantum state tomography offers a comprehensive selection of practical and numerically fast estimation schemes allowing precise estimation of mixed and pure quantum states \cite{Ferrie2014,Granade2017,Quek2021, Qi2017, Barndorff2000, Kalev2015, Zambrano2020, Ahmad2022, vargas2024, Kazim2021, Mondal2023, Farooq2022}. Adaptive quantum state reconstruction strategies have been adopted in various experimental qubit platforms, including photonic systems \cite{Kravtsov2013, Struchalin2018, Struchalin2016, Kravtsov2013, Qi2017, Okamoto2012, Hou2016} and superconducting architectures \cite{Hwang2023}.

\begin{figure}
    \centering
    \includegraphics[width=\linewidth]{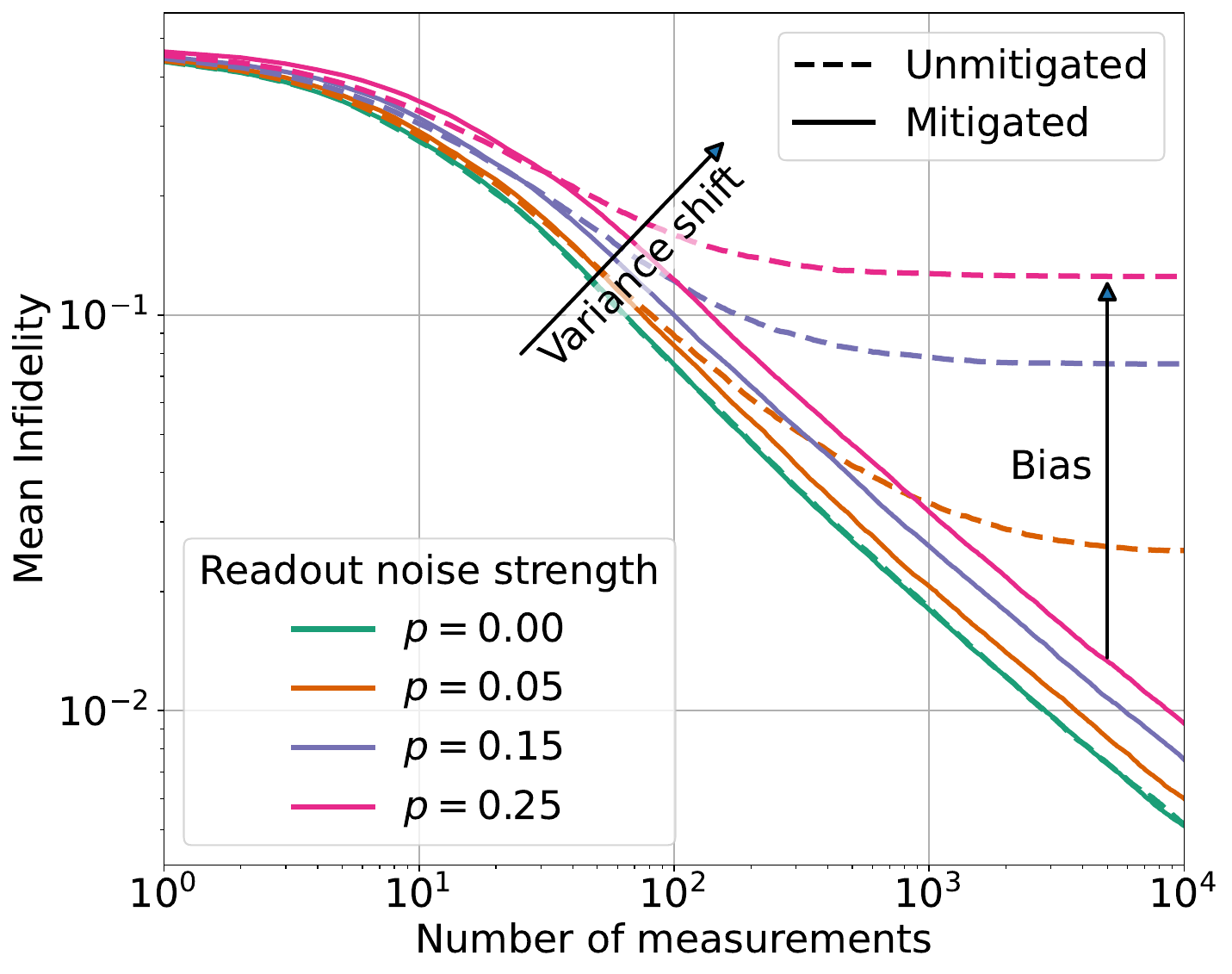}
    \caption{Simulated single-qubit state reconstruction with and without readout-error mitigation for various levels of depolarizing noise $p$. Without readout-error mitigation (dashed lines), the reconstruction infidelity eventually saturates at a bias value set by the distance to the effectively measured noisy state. When exact readout-error mitigation is applied (solid lines) the mean infidelity curves keep decreasing but shift further to the right with increasing readout noise strength. Each curve is averaged over 2000 Haar-random pure states. Both axes are plotted on a logarithmic scale. }
    \label{fig:bias-variance}
\end{figure}

In standard state reconstruction protocols, detectors are typically assumed to be ideal, which makes it straightforward to identify the optimal measurements. In real experiments, however, detectors are noisy, which both restricts the reachable reconstruction infidelity and introduces a noise-induced bias in the estimator. With readout noise, the effective measured state undergoes a transformation through a readout noise channel, denoted $\tilde \rho =\mathcal{E}(\rho)$. Without modifications to the estimator, the infidelity will at some point level off at a biased value, determined by the distance between the target state and the effectively measured state, see Fig.~\ref{fig:bias-variance}. This bias can be removed by integrating readout-error mitigation into the state reconstruction process \cite{Aasen2024, Arrasmith2023, Maciejewski2020}, often by performing detector tomography \cite{Luis1999, Lundeen2008}. In general, error mitigation methods trade estimator bias for increased estimator variance \cite{Cai2023}, where stronger noise typically induces a higher variance penalty, as illustrated in Fig.~\ref{fig:bias-variance}. If the noisy detection process can be characterized exactly, the bias vanishes, and only an asymptotically decaying variance remains. For static, copy-by-copy measurement strategies, pure and close-to-pure states achieve the same asymptotic $1/\sqrt{N}$ infidelity scaling with and without readout noise in the system \cite{Aasen2024, Digiovanni2025}.  While these results may suggest that readout-error-mitigated measurement strategies are robust against readout noise, we find that introducing readout noise fundamentally alters the asymptotic behavior of adaptive measurement strategies. In this work, we demonstrate that even an arbitrarily small amount of readout noise eliminates the asymptotic scaling advantage for adaptive strategies. Furthermore, we observe that the bias–variance trade-off also appears in adaptive measurement strategies. 

Other works have investigated readout noise in the context of optimal quantum state tomography. The decrease in reconstruction infidelity for static measurement schemes caused by noise was investigated in Refs.~\cite{Bogdanov2016, Bogdanov2023}. In Ref.~\cite{IvanovaRohling2023}, the authors introduced a new optimal static measurement approach that considers readout noise modeled as a noisy unitary gate applied before an ideal readout. In Ref.~\cite{Hwang2023} the authors perform noise-aware adaptive measurement optimization with lowered state distinguishability in superconducting circuits, but without any readout-error mitigation. Further works have looked at including noise in optimal adaptive measurement strategies for parameter estimation tasks \cite{Demkowicz2017, Oshio2024}.

In this work, we investigate three different aspects of how noise impacts adaptive measurement strategies. 
First, in Sec.~\ref{sec:method}, we present two different adaptive measurement strategies and derive the conditions required for a measurement strategy to be optimal. Then, in Sec.~\ref{sec:Result_loss_of_optimal_scaling}, we show that imperfect measurements place a nontrivial upper bound on the recoverable infidelity of adaptive measurement strategies for pure and close-to-pure states using Fisher information optimization \cite{Bogdanov2009, Struchalin2018}. This bound causes optimal asymptotic infidelity scaling to be lost with any amount of readout noise in the system.
Secondly, in Sec.~\ref{sec:Result_numerical_transient_behavior}, we verify this claim for single- and two-qubit systems numerically by using Bayesian mean estimation (BME) with minimization of the Bayesian posterior entropy \cite{Huszr2012}. Beyond confirming the loss of optimal scaling, the simulations reveal a bias–variance tradeoff similar to that seen in static measurement schemes. We examine the transient behavior and observe a gradual cross-over from optimal to suboptimal scaling. Despite the fundamental limitations encountered in the asymptotic regime, our results show that adaptive strategies consistently provide a multiplicative improvement in reconstruction accuracy.
Thirdly, in Sec.~\ref{sec:results_realistic_experimental_scenario}, we numerically investigate scenarios in which a limited number of measurements can be used for detector tomography. We find, as expected, that when very few state copies are used for detector tomography compared to state tomography the infidelity curves saturate to a biased value, similar to unmitigated state tomography in Fig.~\ref{fig:bias-variance}. For the type of noise simulated in this work, detector tomography requires at least five times as many state copies as were used for a single state reconstruction, although this factor typically needs to be adjusted individually for each experiment \cite{Digiovanni2025}.

\section{Preliminaries}
\label{sec:preliminaries}
\subsection{Generalized quantum measurements}
A generalized quantum measurement, described by a positive operator-valued measure (POVM), consists of Hermitian operators $\mathbf{M} = \{M_\gamma\}$ that satisfy the following properties
\begin{equation}
    M_\gamma\geq 0, \quad\quad M_\gamma^\dagger = M_\gamma, \quad\quad \text{and} \quad\quad \sum_\gamma M_\gamma = \mathbb{1}.
    \label{eq:POVM_def}
\end{equation}
Each POVM element $M_\gamma$ is associated with an outcome $\gamma$ of a measurement process. The Born rule provides the probabilities for the different possible outcomes when measuring a quantum state $\rho$,
\begin{equation}
\label{eq:Borns_rule}
    \Tr(\rho M_\gamma) = \langle M_\gamma \rangle = p_\gamma.
\end{equation}
One benefit of working with POVMs rather than just projective measurements is that they allow us to capture the process of noise at readout.

In this work, we consider single- and two-qubit states, where we use the Pauli-6 POVM unless specified otherwise,
\begin{equation}
    \mathcal{M}_{\text{Pauli-6}} = \left\{\frac{1}{3}\ket{0_i}\bra{0_i},
    \frac{1}{3}\ket{1_i}\bra{1_i} \right\}\quad  \text{for }i\in\{x,y,z\},
    \label{eq:Pauli-6}
\end{equation}
where $\ket{0_i}$ and $\ket{1_i}$ are the two eigenstates of the corresponding Pauli operators.
The two-qubit Pauli-6 POVM is created by taking the tensor product of all possible combinations of the elements of two single-qubit Pauli-6 POVMs.

\subsection{Readout-error-mitigated state tomography}
In order to reliably reconstruct quantum states despite the presence of readout noise in the system, it is necessary to perform readout-error mitigation. This work uses the protocol presented in Ref.~\cite{Aasen2024} that addresses the mitigation of beyond-classical readout errors in quantum state reconstruction.
The basic outline of the approach is as follows: quantum detector tomography is used as a calibration step, where the actual noisy POVM $\mathcal{M}=\{M_\gamma\}$ implemented by the measurement device is reconstructed \cite{Fiurek2001}. The reconstructed POVM is then integrated into the likelihood function 
\begin{equation}
    \mathcal{L}(\rho) \propto \Pi_\gamma \Tr(\rho M_\gamma)^{n_\gamma},
    \label{eq:likelihood_function}
\end{equation}
where $n_\gamma$ is the number of occurrences of outcome $\gamma$. Using a Bayesian mean estimator \cite{BlumeKohout2010} or a maximum likelihood estimator \cite{Lvovsky2004}, a most probable state can be found that is guaranteed to be physical.

In Secs.~\ref{sec:Result_loss_of_optimal_scaling}
and \ref{sec:Result_numerical_transient_behavior}, we assume that the noisy POVM is determined exactly, i.e. that we have access to the exact measurement operators implemented by the detector for readout-error mitigation. This is necessary to study the asymptotic behavior of adaptive measurement strategies but is not experimentally realistic. To amend this, we include the detector tomography step in the numerical simulations in   Sec.~\ref{sec:results_realistic_experimental_scenario}.

\subsection{Depolarizing noise}
\label{sec:depolarizing_noise}
Depolarizing noise represents a general and typical type of readout noise, and it is the main form of noise that we examine in this work. It is characterized by a lowered state distinguishability, and can be viewed as a redistribution of measurement outcomes. 
The depolarizing channel acting on some operator $A$ in an $n$-qubit Hilbert space with strength $p$ is defined as \cite{Gorini1976, Holevo2012} 
\begin{equation}
\begin{split}
    \mathcal{E}^p_\text{depol}(A) &= (1-p)A +  p\frac{\mathbb{1}}{2^n} \Tr(A) \\ &\stackrel{n=1}{=}\left(1-\frac{3p}{4}\right)A + \frac{p}{4}\sum_{i=1}^3 K_iA K_i^\dagger
    \label{eq:depol_channel}
\end{split}
\end{equation}
where $(K_1,K_2,K_3) = (X, Y, Z)$ are the Pauli matrices. The second form in Eq.~\eqref{eq:depol_channel} is the operator sum form for a single-qubit depolarizing channel. The operator sum form can be generalized to $n$-qubit channels \cite{Magesan2011,Holevo2012}. 

\subsection{Quantum infidelity}
The reconstruction quality metric we use is the quantum infidelity, defined as 
\begin{equation}
    I(\rho,\sigma) = 1-F(\rho,\sigma),
    \label{eq:infidelity}
\end{equation}
where $F(\rho,\sigma)$ is the quantum fidelity \cite{Fuchs1994},
\begin{equation}
    F(\rho,\sigma)=\left[\Tr(\sqrt{\sqrt{\rho}\sigma\sqrt{\rho}})\right]^2,
\end{equation}
 where $\sqrt{\rho}$ should be understood as the square root of the eigenvalues $\lambda_i$ of $\rho$ in an eigendecomposition $\sqrt{\rho}=V\sqrt{D}V^{-1}$, where $\sqrt{D}=\text{diag}(\sqrt{\lambda_1},\sqrt{\lambda_2}, \dots , \sqrt{\lambda_n})$. 
 
The quantum fidelity simplifies in some important cases, most notably for when one state is pure,
\begin{equation}
    F_\text{Pure}\left(\ket{\psi_\rho},\sigma\right)=\bra{\psi_\rho}\sigma \ket{\psi_\rho} = \Tr(\rho \sigma),
\end{equation}
and for qubit states
\begin{equation}
    F(\rho,\sigma)=\Tr(\rho\sigma) + 2\sqrt{\det(\rho)\det(\sigma)}.
    \label{eq:qubitFidelity}
\end{equation}

\subsection{Fisher information matrix }

The Fisher information is formally defined as the variance of the score function
\begin{equation}
    \mathcal{I}(\theta)=\text{Var}(s(\theta)),
\end{equation}
where the score function is
\begin{equation}
    s(\theta)=\frac{\partial \ln(\mathcal{L}(\theta))}{\partial \theta}.
\end{equation}
The Fisher information quantifies how sensitive the likelihood function is to changes in the parameter $\theta$. A high Fisher information indicates that small variations in the measurement data lead to large changes in the inferred parameter, while a low Fisher information implies that the parameter remains relatively insensitive to additional measurements. 
For multiple parameters $\{\theta_i\}$ the Fisher information generalizes to a matrix with two equivalent formulations
\begin{equation}
\begin{split}
\label{eq:Fisher_info_matrix_formulation}
    \left[\mathcal{I}(\theta)\right]_{ij}&=\mathbb{E}\left[  \frac{\partial \ln(\mathcal{L}(\theta))}{\partial \theta_i}   \frac{\partial \ln(\mathcal{L}(\theta))}{\partial \theta_j}  \right]\\ &= -\mathbb{E}\left[  \frac{\partial^2 \ln(\mathcal{L}(\theta))}{\partial \theta_i \partial \theta_j} \right],
\end{split}
\end{equation}
where the $i,j$ entry quantifies the information shared between the parameters $\theta_i$ and $\theta_j$. 

\section{Methods}
\label{sec:method}

\subsection{Defining the estimation problem}
\label{sec:defining_the_estimation_problem}

When estimating quantum states, there are two sources of error that contribute to the infidelity, the bias of the estimator and the estimator variance. The bias is intrinsic to the estimator and remains constant regardless of the number of measurements performed. In contrast, the variance decreases as the number of measurements increases. We aim to investigate how adaptive measurement strategies influence the variance of the estimator, and, except in Sec.~\ref{sec:results_realistic_experimental_scenario}, we assume that any estimator bias has been removed using exact readout-error mitigation.

Most standard measurement strategies have a worst-case infidelity scaling of $\propto 1/\sqrt{N}$ \cite{Mahler2013, Bogdanov2009}. According to the Gill-Massar bound, there is an upper limit to the achievable infidelity scaling of any estimator, which is proportional to $1/N$ \cite{Gill2000}. Consequently, an estimator is considered asymptotically optimal if it obtains the scaling $1/N$ as $N$ approaches infinity, despite the possibility of further improvements to the scaling coefficient being possible.

In most experiments, the target states are pure states, and it would therefore seem natural to use a pure-state estimator. Such a rank-appropriate estimator would recover optimal $1/N$ infidelity scaling without any adaptive measurements at all \cite{Struchalin2018, Xiao2023, Bogdanov2009}. The suboptimal $1/\sqrt{N}$ scaling, as will be elaborated in the following sections, arises from a mismatch between the rank of the estimator and that of the target state, and once the two are matched, even static measurements achieve $1/N$. The question of recovering optimal scaling through adaptivity is thus only meaningful for high-rank mixed-state estimators, which overparametrize the target and place its pure-state component on the boundary of the estimator domain (see Appendix~\ref{app:pure_state_in_mixed_state_estimaton} for an intuitive explanation of the resulting degradation). There are, in addition, two major drawbacks in restricting a state estimator to only pure states. First, the space of pure states is not convex, and it is not guaranteed that the likelihood function is monoidal in this space \cite{BlumeKohout2010}. Secondly, since it is impossible to prepare perfect pure states in any experimental setup, the prepared state is always mixed and would be outside the estimator support. For these reasons our focus will be exclusively on mixed-state estimators. To ensure that the estimators perform optimally on average, the target states will be randomly chosen from a set of Haar-random pure states. In Secs.~\ref{sec:Result_numerical_transient_behavior} and \ref{sec:results_realistic_experimental_scenario} we restrict our numerical analysis to pure states only, ensuring that the $1/\sqrt{N}$ scaling we observe at large $N$ is due to detector noise alone. For close-to-pure but mixed target states, the nonadaptive infidelity would eventually cross over to $1/N$ scaling once $N$ is large enough to resolve the target states' finite distance from the pure-state boundary, which would complicate the comparison.

In summary, our investigation mainly focuses on mixed-state estimators with pure target states using exact readout-error mitigation. The quality measure is the asymptotic scaling of the quantum infidelity in the number of measurements. The quantum infidelity is averaged over Haar-random target states.

\subsection{Adaptive measurement strategies}
\label{sec:adaptive_measurement_strategies}
We examine two kinds of adaptive strategies, those that update the measurement setting a single time and those that update it multiple times. 

\subsubsection{Two-step strategies}
\label{sec:single_step_stratergies}
Single update methods are the more economical of the two approaches, since updating measurement settings can be very slow compared to continuing measurements in the same setting. If the target state is known in advance, the best strategy is to perform a projective measurement in the eigenbasis of the state. This measurement scheme can, in principle, surpass the Gill–Massar bound \cite{Qi2017}. Nonetheless, such an estimator is of limited practical relevance once the target state is already known. A state-agnostic version of this strategy can be created by splitting the estimation process into two steps \cite{Barndorff2000, Mahler2013,Straupe2016, Gill2000}: 
\begin{enumerate}
    \item Perform an initial state estimate $\rho_\text{init}$ based on an informationally complete measurement, such as the $\mathcal{M}_\text{Pauli-6}$ measurements.
    \item Find the eigenstate $\ket{\lambda_1}$ with the largest eigenvalue $\lambda_1$ of $\rho_\text{init}$, and perform the remaining measurements using the generalized measurement $U\mathcal{M}_\text{Pauli-6}U^\dagger$, where $U$ is a unitary rotation matrix that rotates the all-zero qubit state to this eigenstate of $\rho_\text{init}$, $U\ket{0\dots0} = \ket{\lambda_1}$.
\end{enumerate}
The best way to divide the total measurement budget between the two steps has been shown to be to allocate half to each step \cite{Mahler2013}. It was also found to be sensitive to the prior used to initialize the estimator \cite{Straupe2016}. 

Although these methods were found to be optimal, the eigenbasis of multiqubit systems is very often entangled, which typically causes additional overhead to implement experimentally. The underlying mechanism was found to be related to the singular values of the Fisher information matrix \cite{Struchalin2018, Bogdanov2009}, allowing Ref.~\cite{Struchalin2018} to develop a multi-step strategy that uses only the tensor products of measurements on single qubits. This approach is optimal for some multiqubit states, but it is not generally optimal for all states. 
\subsubsection{Multi-step strategies}
\label{sec:multi_step_stratergies}

\begin{figure}
    \centering
    \includegraphics[width=1\linewidth]{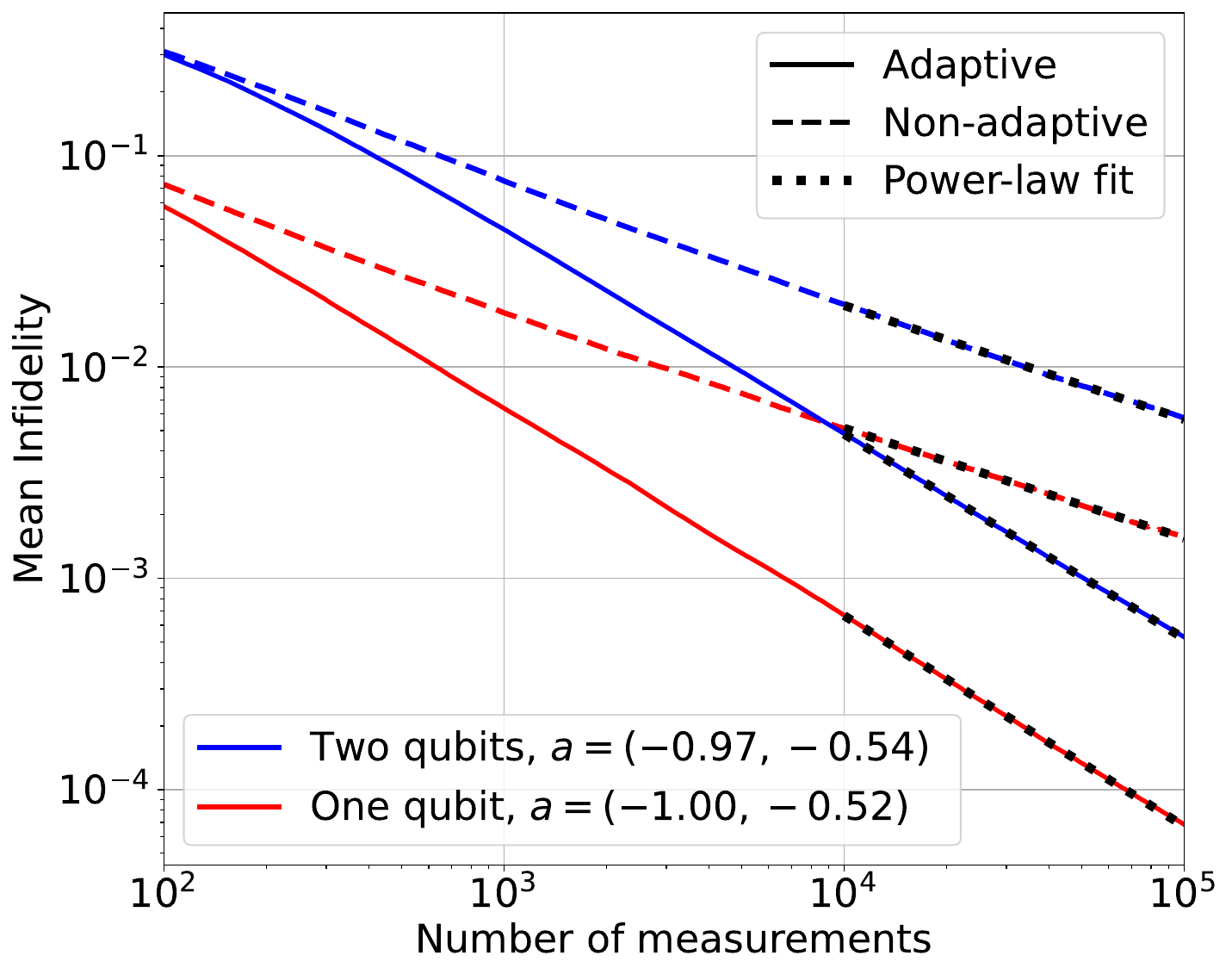}
    \caption{Comparison of mean infidelity scaling between adaptive and nonadaptive Bayesian measurement strategies for single- and two-qubit state reconstruction. A power-law curve $N^a$ is fitted between $10^4$ and $10^5$ measurements, where the scaling exponent $a$ is displayed for (adaptive, nonadaptive) in the legend. The single-qubit nonadaptive approach uses static Pauli-6 measurements. Both adaptive and nonadaptive two-qubit strategies use factorized measurements, i.e. $\mathcal{M} = \mathcal{N}_1\otimes \mathcal{N}_2$. The nonadaptive approach uses static factorized Pauli-6 measurements, i.e. $\mathcal{M} =\mathcal{M}_\text{Pauli-6}\otimes \mathcal{M}_\text{Pauli-6}$. The single- and two-qubit curves are averaged over 2000 and 500 Haar-random states respectively. Both axes are plotted on a logarithmic scale.}
    \label{fig:Adaptive_measurement_stratergy}
\end{figure}

Multi-step strategies are a form of adaptive experimental design, where the estimator continuously updates the measurement strategy based on prior measurement outcomes $D$. A prominent example is Bayesian adaptive tomography, where Bayesian inference is used to incorporate the measurement outcomes into a continuously updated posterior distribution $\pi_f(\rho)$. A new measurement setting $\alpha$ is chosen by maximizing the expected value of a utility function $U(\alpha,\pi_f(\rho))$ based on the current posterior distribution $\pi_f(\rho)$ \cite{Straupe2016}, 
\begin{equation}
    \alpha_n = \argmax_\alpha  \sum_\gamma\int d\rho\, \pi_f(\rho) \Tr(\rho M_\gamma^{\alpha}) U(\alpha, \pi_f(\rho)).
    \label{eq:measurement_optimization}
\end{equation}
There are two common approaches to choosing the utility function, optimizing for a specific parameter, e.g. the infidelity, or maximizing information gain. Refs.~\cite{Huszr2012,Patra2007} demonstrated that using the information gain approach significantly simplifies the computation of the optimization objective in Eq.~\eqref{eq:measurement_optimization}. The optimization can be rewritten as
\begin{equation}
\begin{split}
    \alpha_n =& \argmax_\alpha \bigg( H\left(\int d\rho \, \pi_f(\rho) \Tr(\rho M_\gamma^\alpha)\right)\\ &- \int d\rho \, \pi_f(\rho) H\left(\Tr(\rho M_\gamma^\alpha)\right)\bigg),
    \label{eq:Shannon_optimized_measurement}
    \end{split}
\end{equation}
where $H[p_\gamma] = -\sum_\gamma p_\gamma \log(p_\gamma) $ denotes the Shannon entropy of the discrete probability distribution $p$. The advantage of this reformulation is that it relies solely on the current posterior distribution $\pi_f(\rho)$, and requires only the computation of discrete Shannon entropies. The terms in Eq.~\eqref{eq:Shannon_optimized_measurement} can also be understood intuitively: the first term promotes measurements along the measurement axis with the highest uncertainty (high entropy), while the second term favors measurements where the most likely states have low uncertainty (low entropy). Combined, the optimization seeks the measurement setting $\alpha$ that targets regions of high uncertainty in the outcome distribution while remaining consistent with the most probable candidate states. In Fig.~\ref{fig:Adaptive_measurement_stratergy}, we show that the information-gain strategy achieves optimal asymptotic infidelity scaling for both single- and two-qubit state reconstruction. Notably, in the two-qubit case optimal scaling is reached using only separable measurement operators, whereas the two-step approach can generally not achieve optimal scaling without measurements in entangled bases \cite{Huszr2012, Bogdanov2009, Struchalin2018}.

\subsection{Mechanism behind optimality}
\label{sec:optimal_stratergy_mechanism}
Here we present a summary of the analytical derivation of the conditions that an optimal adaptive measurement strategy must satisfy, formulated in terms of the Fisher information matrix. To maintain conciseness, we focus solely on the primary quantities relevant to the subsequent sections. For a more comprehensive discussion, see Appendix~\ref{App:Optimal_scaling_derivation}.

The basis of this analysis is the mechanism derived in Ref.~\cite{Bogdanov2009}, and refined in Ref.~\cite{Struchalin2018}. 
The starting point of these works is that the state reconstruction infidelity scales, on average, with the singular values $\{\sigma_i\}$ of the Fisher information matrix $\mathcal{I}$ (defined in Eq.~\eqref{eq:Fisher_info_matrix_formulation}),
\begin{equation}
    \langle 1-F\rangle  = \sum_{i=2}^{\nu+1}\frac{1}{\sigma_i},
    \label{eq:infidelity_scaling_singular_values}
\end{equation}
where $\nu$ is the number of degrees of freedom in the estimated quantum state. The sum in Eq.~\eqref{eq:infidelity_scaling_singular_values} starts from $i=2$ because the singular value $\sigma_1$ is related to the normalization of the state and is therefore fixed. Following our estimation problem outlined in Sec.~\ref{sec:defining_the_estimation_problem}, we only consider full-rank estimators, and therefore $\nu = 4^n-1$. The asymptotic scaling of the singular values is proportional to $N$, which means that the expected infidelity scales proportionally to $1/N$ (see Eq.~\eqref{eq:Multinomial_Fisher_matrix} in Appendix \ref{app:Compute_fisher_info_multinomial}). Eq.~\eqref{eq:infidelity_scaling_singular_values} applies only to singular values that are nonvanishing, as in differing from zero beyond statistical fluctuations of the estimator \cite{Bogdanov2009, Struchalin2018, Bagan2006B}. The number of nonvanishing singular values corresponds to the number of degrees of freedom of the target state. For example, if the target state is pure, only $2^n - 1 < \nu$ singular values are nonvanishing. Consequently, the scaling described in Eq.~\eqref{eq:infidelity_scaling_singular_values} holds only for these nonvanishing singular values, while the remaining ones must be treated separately, leading to contributions that scale as $1/\sqrt{N}$ \cite{Struchalin2018, Bagan2006B}. The suboptimal scaling can be understood by considering that, in an overparametrized estimator, certain estimator parameters can lie along the boundary of the allowable parameter space. As a concrete example, consider a single-qubit state represented on the Bloch sphere, with a pure target state. In this case, the true value of the radial component lies exactly on the boundary of its parameter space. We show in Appendix~\ref{app:pure_state_in_mixed_state_estimaton} that, near such a boundary, the infidelity exhibits a square-root degradation in its scaling compared to the case where the true parameter lies in the interior of the parameter space.

The singular values of the Fisher information matrix are related to the eigenvalues of the density matrix. The rows and columns of $\mathcal{I}$ are proportional to the vector \cite{Struchalin2018}
\begin{equation}
\omega^\alpha_\gamma =
\begin{pmatrix}
    \sqrt{\lambda_1}M^\alpha_{\gamma} \ket{\lambda_1}\\
    \sqrt{\lambda_2}M^\alpha_{\gamma} \ket{\lambda_2}\\
    \vdots\\
    \sqrt{\lambda_{2^n}}M^\alpha_{\gamma} \ket{\lambda_{2^n}}\\ 
\end{pmatrix},
\quad \mathcal{I} \propto N\sum_\alpha \sum_\gamma\frac{1}{p^\alpha_\gamma} \omega^\alpha_\gamma (\omega^\alpha_\gamma)^\dagger,
\label{eq:Fisher_zero_rows}
\end{equation}
where $\lambda_i$ are the eigenvalues of the target state with the corresponding eigenstates $\ket{\lambda_i}$, and $\gamma$ labels the outcome of measurement setting $\alpha$ with respective probability $p^\alpha_\gamma$,
 \begin{equation}
     p^\alpha_\gamma = \frac{1}{\tau_\gamma^\alpha} \sum_i  || \sqrt{\lambda_i}M^\alpha_{\gamma} \ket{\lambda_i}||_2^2,
 \end{equation}
where $\tau_\gamma^\alpha = \Tr(M_\gamma ^\alpha)$ and $||\ket{\psi}||_2 = \sqrt{\bra{\psi}\ket{\psi}}$. From Eq.~\eqref{eq:Fisher_zero_rows}, we see that if any eigenvalues $\lambda_i$ are small (compared to the statistical uncertainty of the estimator), entire rows and columns of $\mathcal{I}$ become suppressed. As a result, the Fisher information matrix effectively behaves as rank deficient, reflecting a mismatch between the degrees of freedom of the estimator and those of the target state. To recover optimal scaling, one should choose a measurement strategy that makes the estimator effectively behave as if it were full rank.

This can be achieved by amplifying the suppressed rows and columns of $\mathcal{I}$ through the factor $1/p^\alpha_\gamma$. In particular, one can choose a measurement setting $\mathcal{M}^\alpha$ containing elements $M^\alpha_\gamma$ that are orthogonal to the eigenvectors associated with the nonvanishing eigenvalues, i.e., $M^\alpha_\gamma \ket{\lambda_i} = 0$ for all $\lambda_i \gg 0$. Such measurements isolate the components of $\omega^\alpha_\gamma (\omega^\alpha_\gamma)^\dagger$ corresponding to the small eigenvalues. The associated probability $p^\alpha_\gamma$ is itself small and proportional to the suppressed entries of $\omega^\alpha_\gamma (\omega^\alpha_\gamma)^\dagger$, 
\begin{equation}
p_\gamma ^\alpha \propto 
    || \sqrt{\lambda_i}M^\alpha_{\gamma} \ket{\lambda_i}||^2_2 \approx  \, || \sqrt{\lambda_j}M^\alpha_{\gamma} \ket{\lambda_j}||^2_2,
    \label{eq:equal_g_factor}
\end{equation}
for all $\lambda_i\approx \lambda_j \approx 0$. Consequently, the prefactor $1/p^\alpha_\gamma$ compensates for this suppression, producing a new nonvanishing singular value in the Fisher information matrix. This procedure can be repeated for multiple measurement operators $M_\gamma^\alpha$ until the Fisher information matrix becomes full rank.

We can summarize the above discussion succinctly. To recover optimal asymptotic scaling, the measurement strategy must ensure that all singular values of the Fisher information matrix are nonvanishing. This is achieved by selecting measurement operators that are orthogonal to the eigenvectors associated with nonvanishing eigenvalues of the target state. In other words, one needs to find measurement operators of the form
 \begin{equation}
 \begin{split}
M^\alpha_\gamma=\tau^\alpha_\gamma\ket{\psi^\alpha_\gamma}\bra{\psi^\alpha_\gamma} &\text{ such that } \bra{\psi^\alpha_\gamma}\ket{\lambda_k} = 0\\ \text{ for all } &k \text{ with } \lambda_k\neq0,
 \end{split}
\end{equation}
 and the number of such measurement operators required for $n$-qubit states is given by \cite{Struchalin2018}
\begin{equation}
N_\mathcal{M} =
4^n - 1 - \big(R2^{n+1} - R^2 - 1\big),
\end{equation}
where $R$ is the number of nonvanishing eigenvalues in the target state.

The two adaptive approaches discussed in Secs.~\ref{sec:single_step_stratergies} and \ref{sec:multi_step_stratergies}  achieve optimal infidelity scaling in slightly different ways. The two-step adaptive strategy achieves equalization of the terms in Eq.~\eqref{eq:equal_g_factor} by realigning the measurement operators to be orthogonal to the nonvanishing eigenstates of an intermediate state estimate $\rho_\text{init}$ \cite{Struchalin2018, Bogdanov2009}. These intermediate eigenstates are often entangled in the multiqubit case, which means that optimal scaling cannot always be achieved with separable measurements with a single measurement setting. However, optimal scaling can be achieved using multiple settings \cite{Struchalin2018}. Multi-step adaptive Bayesian strategies do not explicitly optimize for orthogonal measurement operators, but rather continuously update the measurement strategy based on the uncertainty of the posterior distribution $\pi_f(\rho)$ \cite{Struchalin2016, Huszr2012}. The different measurement settings are amplifying the small singular values in a comparable way, within the current estimator uncertainty, and as additional measurements are performed, the measurement operators are progressively adjusted to become more orthogonal to the eigenstates corresponding to nonvanishing eigenvalues.

\section{Loss of optimal asymptotic scaling}
\label{sec:Result_loss_of_optimal_scaling}
To investigate how adaptive measurement strategies perform in the presence of detector noise, we examine how the mechanism described in Sec.~\ref{sec:optimal_stratergy_mechanism} is affected when depolarizing readout noise is introduced. Depolarizing noise (defined in Sec.~\ref{sec:depolarizing_noise}) serves as a natural stand-in for experimental noise, as most readout imperfections manifest as a statistical redistribution of measurement outcomes, which can be modeled as a type of depolarizing channel. In addition to being an effective noise model commonly observed in experiments \cite{Maciejewski2020, Smith2021}, Pauli twirling can transform many nontrivial noise channels into effective depolarizing channels \cite{Cai2023}.

With this simple assumption, it is possible to put an upper bound on the recoverable asymptotic infidelity for adaptive measurement strategies. Using the readout-error-mitigated likelihood function in Eq.~\eqref{eq:likelihood_function} and rederiving the vector $\omega^\alpha_\gamma$, the ideal measurement operators in Eq.~\eqref{eq:Fisher_zero_rows} are replaced by their depolarized counterparts
\begin{equation}
    M^\alpha_\gamma \rightarrow \mathcal{E}^p_\text{depol}(M^\alpha_\gamma) = \tilde M^\alpha_\gamma,
\end{equation}
where we have used the depolarizing channel from Eq.~\eqref{eq:depol_channel}. The depolarized measurement operator $\tilde{M}^\alpha_\gamma$ contains an identity component $\mathbb{1}p\tau^\alpha_\gamma/2^n$, which contributes a state-independent baseline to every outcome probability. Consequently, expanding $p^\alpha_\gamma = \text{Tr}(\rho \tilde{M}^\alpha_\gamma)$ and using $\lambda_i|\langle\psi^\alpha_\gamma|\lambda_i\rangle|^2 \geq 0$, one obtains the lower bound
\begin{equation}
    p^\alpha_\gamma  \geq \tau^\alpha_\gamma\frac{p}{2^n},
    \label{eq:lower_noise_bound_probabilities}
\end{equation}
which limits how much the factor $1/p^\alpha_\gamma$ in Eq.~\eqref{eq:Fisher_zero_rows} can amplify the suppressed rows and columns in the Fisher information matrix. Note that choosing $\tau^\alpha_\gamma\rightarrow 0$ does not help since $\tfrac{1}{p^\alpha_\gamma} \omega^\alpha_\gamma (\omega^\alpha_\gamma)^\dagger \propto \tau^\alpha_\gamma$. It follows that no multi-step adaptive protocol can recover the optimal asymptotic infidelity scaling in the presence of depolarizing readout noise. This constitutes the main analytical result of this work, with the full derivation of Eq.~\eqref{eq:lower_noise_bound_probabilities} provided in Appendix~\ref{App:Optimal_scaling_derivation}. Importantly, this conclusion concerns only the asymptotic regime under exact readout-error mitigation and does not exclude potential advantages of adaptive measurement strategies in transient, finite-sample regimes, which we discuss next. Note that this bound concerns full-rank estimators. A rank-appropriate estimator would not require amplification of suppressed singular values and retains $1/N$ scaling even with readout noise (see Sec.~\ref{sec:defining_the_estimation_problem}).

It is furthermore clear that although we only consider depolarizing noise here, the results remain valid for other classes of noise channel. In particular, all channels that can be decomposed into a depolarizing channel plus another nontrivial channel will have a nontrivial lower bound given by the depolarizing channel. Other common noise channels such as bit-flip, phase-flip, and qubit decay only have two or a single state which remains unaffected by the noise channel, and their contributions under the Haar-measure vanish.  

\section{Numerical transient behavior}
\label{sec:Result_numerical_transient_behavior}
While the result presented in Sec.~\ref{sec:Result_loss_of_optimal_scaling} is conclusive for both two-step and multi-step adaptive methods, we continue to explore the transient behavior of the multi-step method here. For these simulations, we use Bayesian mean estimation. Implementation details are given in Appendix~\ref{app:Implementation_BME}. For adaptive measurement optimization we use the multi-step information gain maximization described in Sec.~\ref{sec:multi_step_stratergies}. Specifically, we use the reformulated objective function in Eq.~\eqref{eq:Shannon_optimized_measurement}. 

\subsubsection{Single- and two-qubit transient infidelity}

\begin{figure*}
 \centering
 \includegraphics[width=0.95\linewidth]{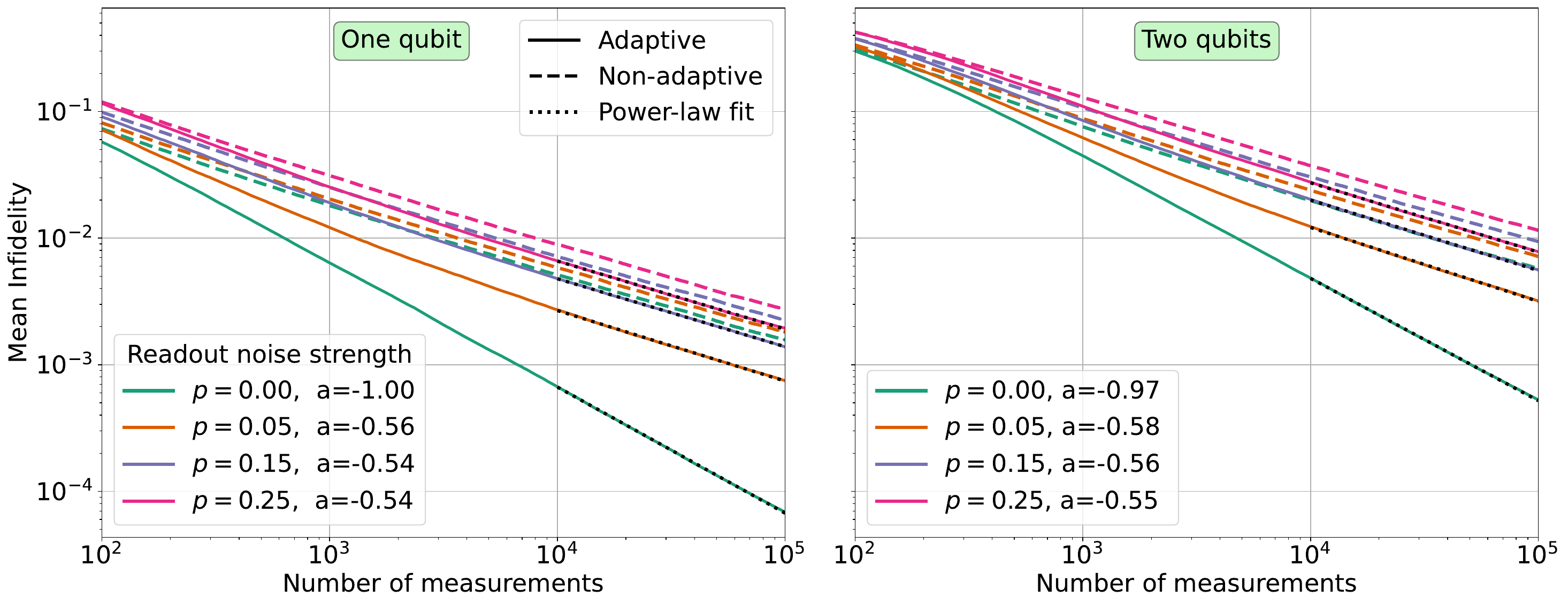}
 \caption{Single- and two-qubit mean infidelities with different amounts of depolarizing readout noise. A power-law $N^a$ is fitted to all adaptive curves within the range of $10^4$ to $10^5$ measurements. The resulting scaling exponents $a$ are displayed in the legend. Both axes are plotted on a logarithmic scale.  \textbf{Left:} Average over 2000 Haar-random single-qubit states. \textbf{Right:} Average over 500 Haar-random two-qubit states.}
 \label{fig:infidelity_comparison}
\end{figure*}

In Fig.~\ref{fig:infidelity_comparison} we show the mean infidelity curves for multi-step adaptive Bayesian mean estimates under different strengths of depolarizing readout noise. In both the single- and two-qubit case, the noise-free scenario exhibits the expected optimal $\approx 1/N$ scaling, while even small amounts of depolarizing readout noise degrade the scaling exponents. A further increase in readout noise strength leads to an increased bias–variance shift associated with performing readout-error mitigation at higher noise levels but leaves the scaling behavior unchanged. For mixed target states, no deterioration is found by introducing readout noise, see Appendix~\ref{app:mixed_target_states}.

\begin{figure}
 \centering
 \includegraphics[width=\linewidth]{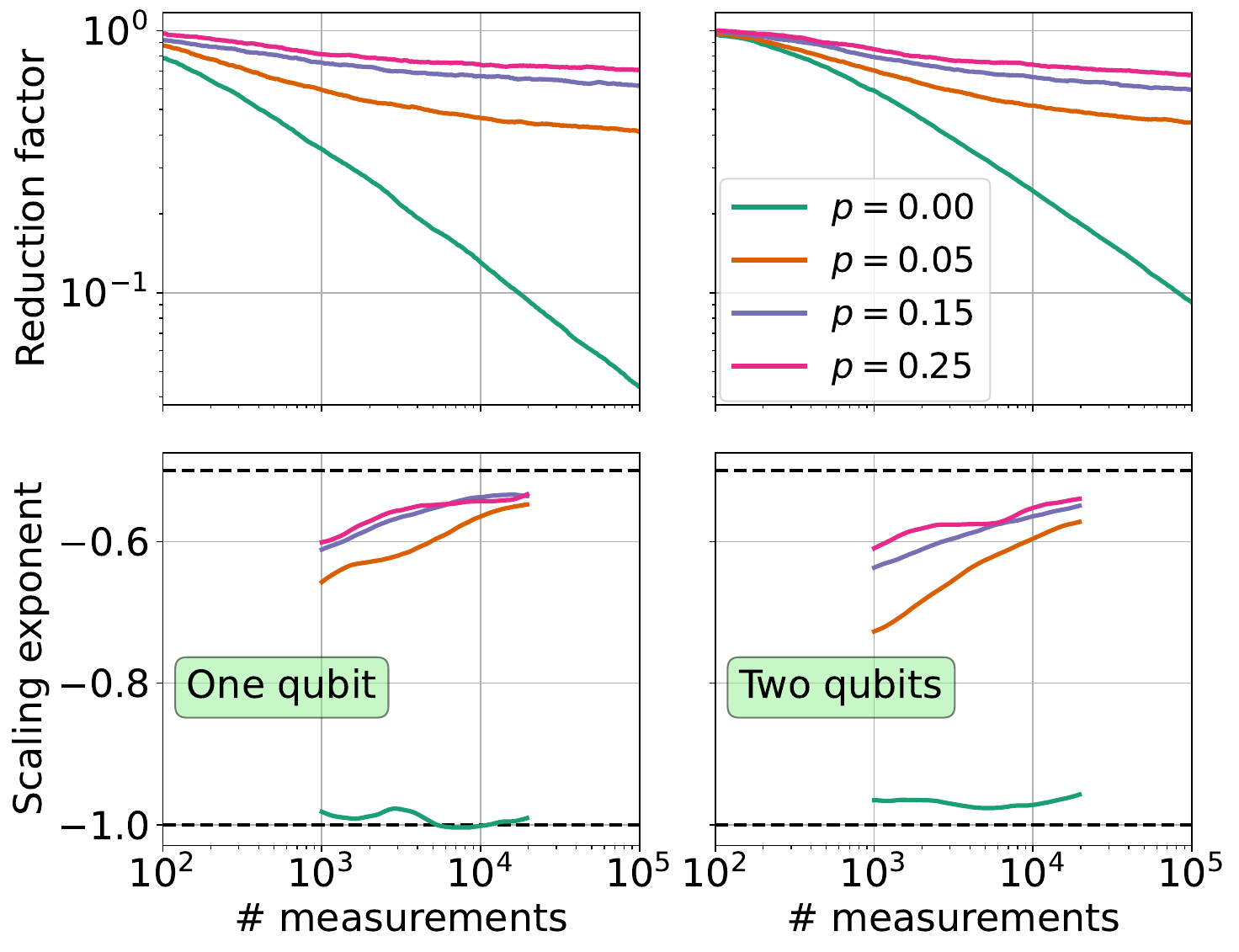}
 \caption{\textbf{Top:} Reduction factor $I_{\text{ada}}/I_{\text{non-ada}}$ between adaptive and nonadaptive method. The zero-noise case continuously decreases while noisy cases level off towards a constant reduction factor. The stronger the readout noise is, the larger the improvement factor becomes. Both axes are plotted on a logarithmic scale. \textbf{Bottom:} Rolling power-law fit  $I \propto N^{a}$ applied to the adaptive curves in Fig.~\ref{fig:infidelity_comparison}, where $a$ denotes the scaling exponent. The rolling power-law fit is applied in the range $[x,5x]$, where $x$ is the measurement number. Only ranges that can be filled are plotted. The black dashed horizontal lines are for guidance, showing expected scaling for adaptive and nonadaptive curves. The asymptotic scaling decays faster for stronger readout noise. The $x$-axis is plotted on a logarithmic scale. }
 \label{fig:Transient_behaviour}
\end{figure}

A generic feature of the low-noise cases is that they initially follow the ideal asymptotic scaling and then gradually deteriorate as the estimation accuracy improves, eventually settling into the suboptimal $1/\sqrt{N}$ scaling characteristic of nonadaptive approaches. Notably, a residual accuracy advantage of adaptive strategies persists even in the presence of depolarizing readout noise.
In Fig.~\ref{fig:Transient_behaviour}, we show the reduction factor in infidelity between adaptive and nonadaptive strategies, $I_\text{ada}/I_\text{nonada
},$ and the scaling exponent extracted from a power-law fit in a rolling window of the number of measurements for the adaptive measurement strategies. The reduction factor levels off for the noisy cases while the noiseless case keeps decreasing the reduction factor. The value at which the reduction factor levels off depends on the strength of the readout noise. For low readout noise, a significant accuracy gain can be obtained by using adaptive measurement strategies. This indicates that for well-calibrated systems, practical benefits can be expected from using adaptive measurement strategies. The rolling fit of the scaling exponent for both single- and two-qubits decreases towards $a=-0.5$  as the number of measurements increases and decreases earlier for higher readout noise. This indicates that the loss of optimal scaling is related to both the accuracy of the estimator and the information-scrambling effect of readout noise. Interquartile ranges for Fig.~\ref{fig:infidelity_comparison} can be found in Appendix~\ref{app:interquartile_range}, and estimator uncertainty is discussed
in Appendix~\ref{app:Estimator_uncertainty}.

\subsubsection{Infidelity distribution at a given number of measurements }
\begin{figure}
 \centering
 \includegraphics[width=\linewidth]{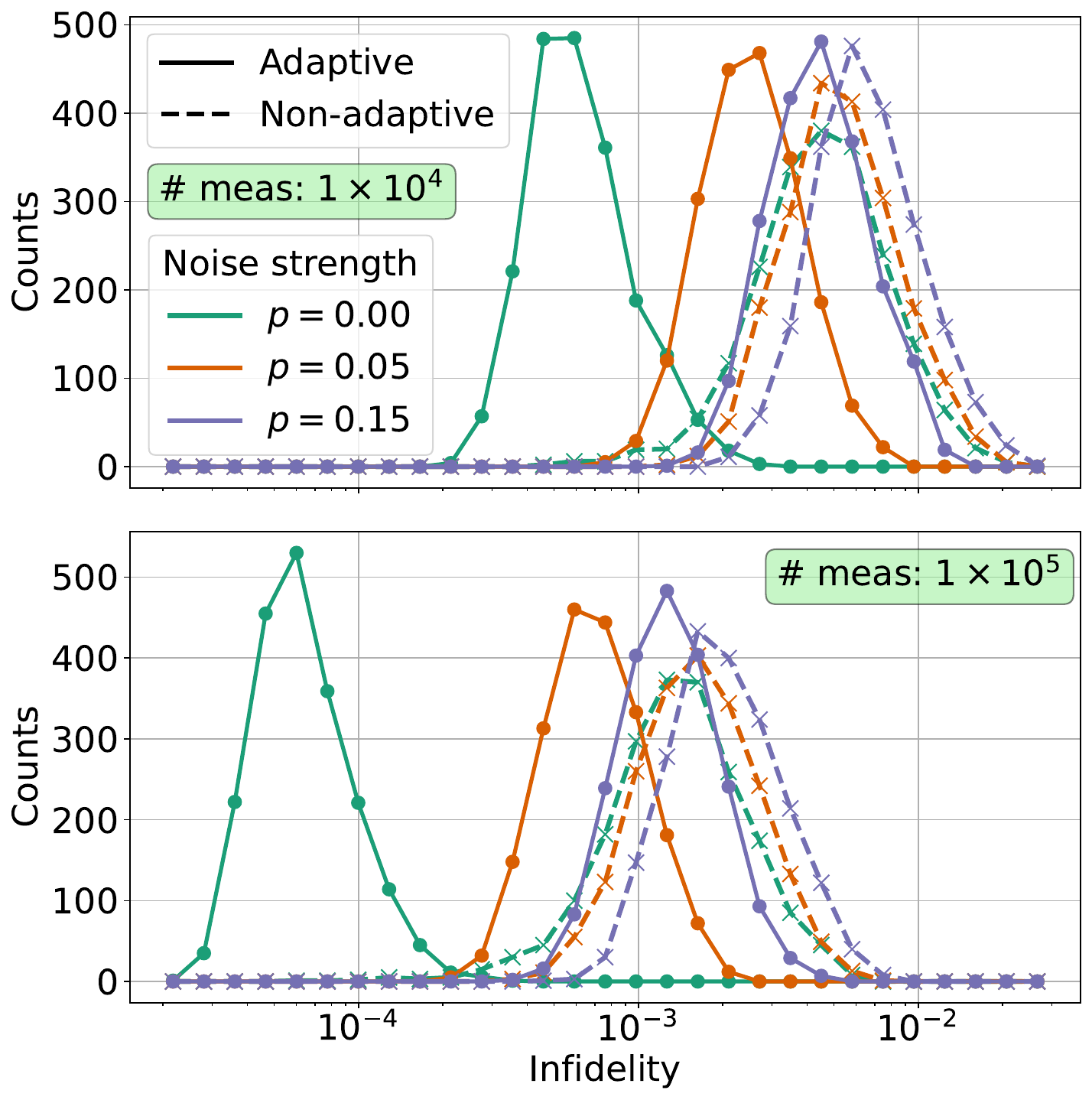}
 \caption{Distribution of the infidelities reached at a given number of measurements for single-qubit curves in Fig.~\ref{fig:infidelity_comparison}. The noiseless adaptive cases separate from the remaining curves which stay relatively close to each other. The $x$-axis is plotted on a logarithmic scale.}
 \label{fig:infidelity_at_shot_threshold}
\end{figure}

In Fig.~\ref{fig:infidelity_at_shot_threshold}, we present the binned distribution of the infidelity reached for a fixed number of measurements, which corresponds to vertical cross sections of all curves in Fig.~\ref{fig:infidelity_comparison}. We see that the adaptive distributions are slightly narrower than their nonadaptive counterparts. The noiseless adaptive case separates from the rest, while the relative distances between the noisy adaptive and nonadaptive distribution peaks remain the same. A similar binned distribution for the number of state copies required to reach a specified infidelity is presented in Appendix~\ref{app:shots_to_reach_infidelity}.

\section{Finite-sample readout-error mitigation}
\label{sec:results_realistic_experimental_scenario}
In the previous sections, we have assumed that the noisy measurement operators $M_\gamma$ required for the readout-error-mitigated state estimator were known exactly. Here, we turn to a scenario in which one must first perform quantum detector tomography with a finite number of samples to obtain estimates of these noisy measurement operators. Similarly to how state tomography reconstructs an unknown state with known measurement operators, detector tomography reconstructs unknown measurement operators with known calibration states \cite{Luis1999, Lundeen2008}. Our calibration states are the tetrahedron states
\begin{equation}
    \begin{split}
        \ket{\psi_1} &= \ket{0},\\
    \ket{\psi_2} &= \frac{1}{\sqrt{3}} \ket{0} + \sqrt{\frac{2}{3}} \ket{1},\\
    \ket{\psi_3} &= \frac{1}{\sqrt{3}} \ket{0} + \sqrt{\frac{2}{3}} e^{i \tfrac{2\pi}{3}}\ket{1},\\
    \ket{\psi_4} &= \frac{1}{\sqrt{3}} \ket{0} + \sqrt{\frac{2}{3}} e^{i \tfrac{4\pi}{3}}\ket{1}.\\
    \label{eq:Tetrahedron_states}
\end{split}    
\end{equation}
Each tetrahedron state is prepared the same number of times for each set of measurement operators. The measurement operators are reconstructed by solving the linear set of equations using the maximum-likelihood estimator \cite{Fiurek2001}
\begin{equation}
f_{s\gamma} = \Tr(\rho_s M_\gamma),
\end{equation}
where $f_{s\gamma}=n_{s\gamma}/N$ are the observed frequencies for outcome $\gamma$ and calibration states $\rho_s = \ket{\psi_s}\bra{\psi_s}$, with $s \in \{1,2,3,4\}$. To compare between adaptive and nonadaptive measurement strategies, we only calibrate the diagonal entries of the measurement operators in the computational basis. This corresponds to only calibrating computational basis measurements, and discarding any phase information. This means that the Pauli-6 POVM used for the nonadaptive measurement strategy is simulated as three separate computational basis measurements with perfect unitary rotations beforehand. Similarly for the adaptive strategy, we assume that each requested setting can be implemented by perfect unitary rotations followed by a depolarized computational basis measurement. 

\begin{figure}
 \centering
 \includegraphics[width=\linewidth]{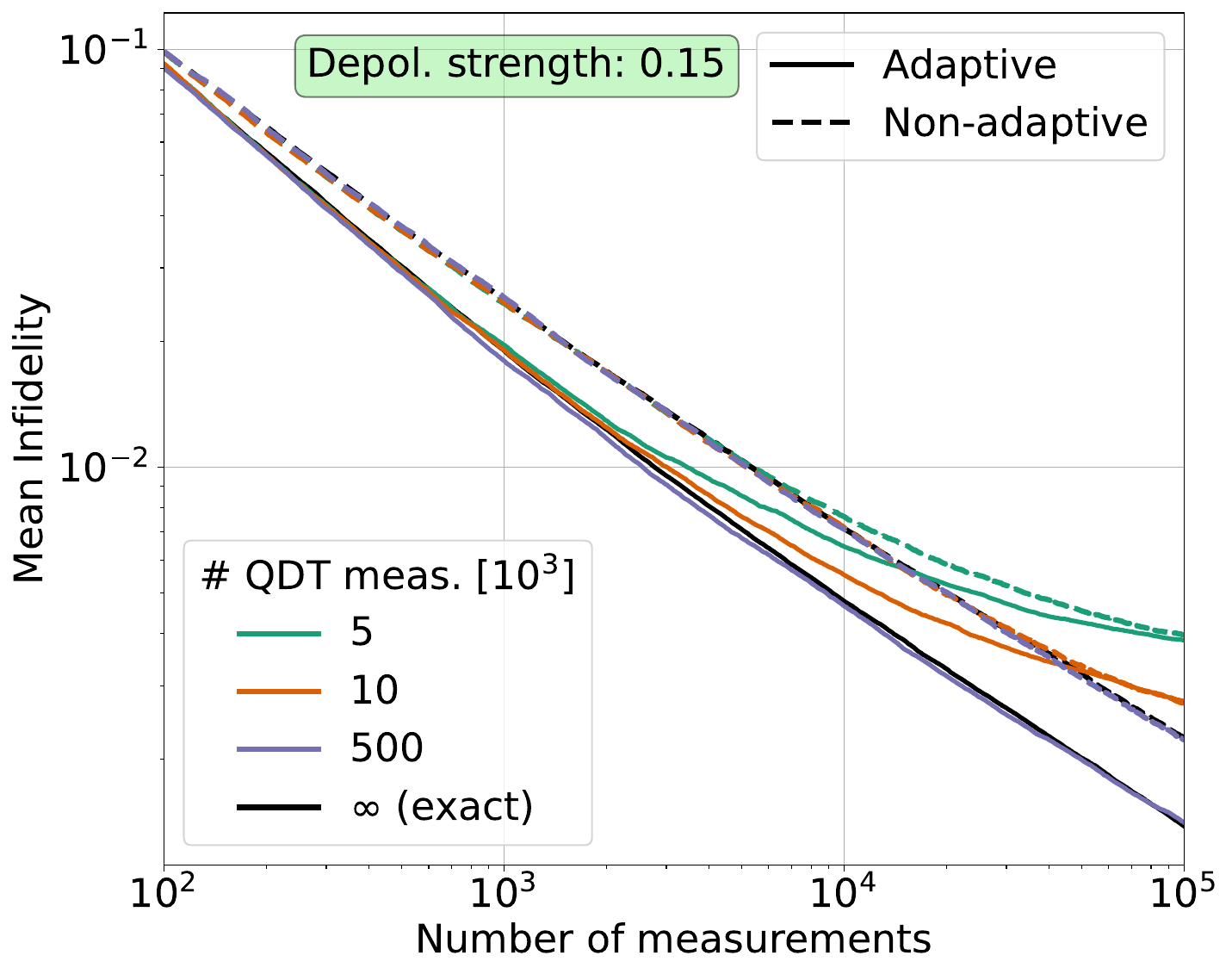}
 \caption{ Mean infidelity curves with different number of state copies used for quantum detector tomography (QDT).  The simulated measurement operators are depolarized with strength $p=0.15$. The curves are averaged over 1000 Haar-random target states, where for each target state detector tomography has been independently performed.  The black curves use the exact depolarized measurement operators, and are equivalent to the curves used in Fig.~\ref{fig:infidelity_comparison}. Both axes are plotted on
a logarithmic scale.}
 \label{fig:qdt_main}
\end{figure}

In Fig.~\ref{fig:qdt_main} we show simulations of state reconstruction where a varying number of state copies have been used for detector tomography. We see that when detector tomography uses fewer state copies, the reconstruction accuracy saturates to a finite value. This behavior is expected because imperfect detector calibration introduces a bias in the estimator. To investigate whether over- and underestimated noise contribute differently to the averaged infidelity, we performed additional simulations in which the reconstructed measurement operators have a constant depolarizing offset relative to the POVM used to generate the data, which is presented in Appendix~\ref{app:qdt_appenidx}. We observed that underestimated noise dominates for pure states and reproduces the characteristics shown in Fig.~\ref{fig:qdt_main}, while overestimated noise yields better reconstruction accuracy due to overestimation of pure states. It is worth noting that when very few state copies are used for detector tomography (here less than state copies used for state tomography), the estimator does not converge properly and should only be used to indicate infidelity saturation, see Appendix~\ref{app:Estimator_uncertainty} for more information.

Fig.~\ref{fig:qdt_main} indicates that, to make sure that accuracy of the reconstructed measurement operators does not limit the quality of the state reconstruction, one should allocate at least five times as many state copies to detector tomography as to state tomography. Depending on the exact structure of the readout noise, one might need more or less state copies and the amount should be tuned for each experiment \cite{Digiovanni2025}. For the detector tomography simulations, 2000 states were used in the particle swarm to have better convergence for the Bayesian estimation, see the Appendix~\ref{app:Implementation_BME} for more information.

\section{Conclusion and outlook}
We investigated how detector noise affects adaptive measurement strategies for quantum state reconstruction with exact readout-error mitigation. We focused specifically on pure- and close-to-pure-state reconstruction in mixed-state estimators, which is a particularly challenging estimation problem \cite{Struchalin2016, BlumeKohout2010}. Adaptive measurement strategies have the potential to improve the asymptotic scaling of reconstruction infidelity from $1/\sqrt{N}$ to $1/N$. We showed that with any amount of depolarizing readout noise, the asymptotic benefit gained from performing adaptive measurements vanishes in the nontrivial regime where the detector noise is characterized exactly, thereby removing any noise-induced bias. This breakdown is derived analytically using Fisher information optimization for the estimator, showing that a sub-optimal contribution to the averaged infidelity cannot be removed when readout noise is present. Furthermore, we analyzed numerically the transient regimes with exact readout-error mitigation and found a gradual decrease of the scaling exponent as the number of measurements increases. Our results show that adaptive measurement strategies maintain a constant-factor gain $I_{\text{ada}}/I_{\text{non-ada}}$, while both $I_{\text{ada/non-ada}}\propto 1/\sqrt{N}$. Both the constant factor gain and the transient scaling improve as readout noise decreases. This indicates that a practical advantage with adaptive measurement strategies can be gained with well-calibrated experiments. We investigated numerically the performance in scenarios where a finite number of state copies are used for detector characterization, by performing quantum detector tomography before performing adaptive quantum state tomography. As expected, not using enough state copies to calibrate the measurement operators causes the state reconstruction accuracy to saturate. For the depolarizing noise simulated in this work, we found that one should use at least five times the number of state copies used for a single state reconstruction for detector tomography. 

These findings constrain how useful adaptive strategies can be in practical state reconstruction tasks, since noise can never be entirely removed on a hardware level. Nevertheless, we find that if the readout noise is low, a significant constant-factor improvement can be gained by using adaptive strategies. Our investigation focused solely on depolarizing noise, which is sufficiently general to cover most relevant cases \cite{Maciejewski2020, Smith2021, Cai2023}. However, it is possible that device-specific noise models might reveal methods capable of maintaining better asymptotic scaling despite having readout noise. It is still unclear whether these results extend to more general measurement strategies, such as collective measurements that use quantum communication \cite{Bagan2006B}. The next step would be to verify this property experimentally, as current adaptive experiments do not perform readout-error mitigation, allowing only the extraction of transient scaling \cite{Kravtsov2013,Struchalin2018, Struchalin2016}. It would be interesting to see if the implementation of noise-conscious measurement strategies would allow some recovery of the benefit of adaptive measurements \cite{IvanovaRohling2023}. Furthermore, it would be interesting to investigate whether it is possible to derive a direct connection between the Fisher information adaptive strategy \cite{Bogdanov2009,Struchalin2018} and the Bayesian utility function approach \cite{Huszr2012, Struchalin2016}. Another open question is the relationship between the transient scaling exponents and readout noise strength with increasing number of measurements. It would be interesting to investigate the dependence of the improvement factor on the detector noise strength and target state purity. Similar factor improvements are observed for mixed target states (Appendix ~\ref{app:mixed_target_states}), indicating that this is a general feature of adaptive measurement strategies for mixed states. A recent work \cite{xiao2025} introduced a unified framework for performing optimal adaptive measurement strategies for state, process, and detector tomography, with associated generalized infidelity metrics for process and detector tomography, with a different approach than Refs.~\cite{Bagan2006B, Struchalin2016}. It would be interesting to investigate whether a similar approach as the one presented in this manuscript would generalize to process and detector tomography in the unified framework of Ref.~\cite{xiao2025}.  

\section*{Code and data availability}
The code used to generate and analyze the results presented can be found, together with a tutorial notebook and the compiled data set, in Ref.~\cite{CREMST_CODE}. The raw dataset used for this project is available in Ref.~\cite{CREMST_DATA}. 

\section*{Author Contributions}
A.S.A. developed the protocol and software under the supervision of M.G.  A.S.A. and M.G. wrote the manuscript.

\section*{Acknowledgments}
The authors thank Etienne Stock for helping to derive the mechanism behind optimality. 
Some python packages were instrumental in this project, Numpy \cite{harris2020}, Scipy \cite{Virtanen2020}, joblib \cite{joblib} and Matplotlib \cite{Hunter2007}. We acknowledge support by the Federal Ministry of Research, Technology and Space (BMFTR) under project BeRyQC.
This research is supported by funding from the German Research Foundation (DFG) under the project identifiers Grant No.\ 398816777-SFB 1375 (NOA) and
Grant No.\ 550495627-FOR 5919 (MLCQS).

\appendix

\section{Pure states in mixed-state estimation}
\label{app:pure_state_in_mixed_state_estimaton}
Estimating pure states proves to be more challenging than estimating mixed states when using mixed-state estimators. Given the central role that pure states play in quantum technologies and information theory, it is essential to understand their distinct behavior. 

We illustrate this issue by considering two single-qubit states 
\begin{equation}
    \rho = \tfrac{1}{2}(\mathbb{1} + \Vec{r} \cdot \Vec{\sigma}) \quad \text{and} \quad \tau = \tfrac{1}{2}(\mathbb{1} + \Vec{t} \cdot \Vec{\sigma}),
\end{equation}
where $\tau$ denotes the target state and $\rho$ denotes the estimated state where each vector component is estimated using $N$ samples. In the Bloch representation, the infidelity can be rewritten as 
\begin{equation}
    I(\rho,\tau)=\frac{1}{2}\left(1-\Vec{r}\cdot\Vec{t} - \sqrt{1-r^2}\sqrt{1-t^2}\right),
    \label{eq:BlochFidelity}
\end{equation}
where we have used the identity
\begin{equation}
    \det(\Vec{r}\cdot\Vec{\sigma})=-r^2,
\end{equation}
where $r=|\Vec{r}|$ and $t=|\Vec{t}|$.
For simplicity, we will assume that the two Bloch vectors are coaxial, $\Vec{r} = r \hat n$ and \mbox{$\Vec{t}= t \hat n$}, for some unit vector $\hat n$. The variance of the radial component is $\text{Var}(r) = \sigma_r^2/N$, where $\sigma_r^2$ is the variance of individual measurements of the Bloch-vector components.

First, we will consider the case where the target state is mixed, $t \sim r < 1$ and $\sigma_r^2/N\ll1-t$. We can therefore relate the lengths of the two vectors through 
\begin{equation}
r = t + \delta,
\end{equation}
where $\delta \propto 1/\sqrt{N}$ is a random variable drawn from a Gaussian distribution, $\delta \sim  \mathcal{N}(0, \sigma_r^2/N)$. Using the fidelity in Eq.~\eqref{eq:BlochFidelity} and the expansion 
\begin{equation}
    \sqrt{1 - (t+\delta)^2} = \sqrt{1-t^2} -\frac{t\delta}{\sqrt{1-t^2}} - \frac{\delta^2}{2(1-t^2)^{3/2}} + O(\delta^3),
\end{equation}
we get
\begin{equation}
I(\rho, \tau)  = \frac{\delta^2}{4 \sqrt{1-t^2}} + O(\delta^3).
\end{equation}
The infidelity therefore scales as $1/N$ for strongly mixed states.

Second, consider the case where the target state is nearly pure and cannot be distinguished from pure states within the estimator’s uncertainty, $1 - t \ll \delta \ll 1$, and we constrain the estimated Bloch vector to remain physical, $r \leq 1$. We can write
\begin{equation}
    r = 1-\xi,
\end{equation}
where $\xi\propto1/\sqrt{N}$ is a random variable from some clipped Gaussian distribution. The infidelity simplifies and we get
\begin{equation}
    I(\rho, \tau) \approx \xi/2.
\end{equation}
Thus, we conclude that when estimating pure states, or states closer to the surface of the Bloch sphere than the statistical error interval of the estimator, the scaling is expected to be square-root worse compared to mixed states, specifically
\begin{equation}
    I(\rho,\sigma)\propto
    \begin{cases}
        \tfrac{1}{\sqrt{N}} & \text{for pure states },\\
        \tfrac{1}{N} &\text{for mixed states}.
    \end{cases}
\end{equation}

\section{Scaling behavior of rank deficient estimators}
\label{App:Optimal_scaling_derivation}

In this section, we provide details of the connecting steps between the results presented in Sec.~\ref{sec:Result_loss_of_optimal_scaling} and the original work, Ref.~\cite{Bogdanov2009}, and the refined explanation found in Ref.~\cite{Struchalin2018}.

Our goal is to identify conditions on the generalized measurement $\mathcal{M}=\{M_\gamma\}$, under which the singular values $\sigma_i$ of the Fisher information matrix scale optimally. We have dropped the $\alpha$ superscript for convenience;  it can always be reabsorbed into $\gamma$. For singular values that are nonvanishing relative to the uncertainty of the estimator, the reconstruction error scales inversely to $\sigma_i$. In this regime, the mean infidelity is given by Eq.~\eqref{eq:infidelity_scaling_singular_values}, which we restate here for convenience,
\begin{equation}
        \langle 1-F\rangle  = \sum_{i=2}^{\nu+1}\frac{1}{\sigma_i}.
     \label{eq:app_mean_inf_singular_values}
\end{equation}
If some singular values vanish, they correspond to directions at the boundary of the parameter space and must be treated separately. In the limit $\sigma_i \to 0$, such singular values generically scale as $1/\sqrt{N}$ rather than $N$, reflecting the boundary nature of the estimation problem \cite{Struchalin2018}, see also Appendix~\ref{app:pure_state_in_mixed_state_estimaton}.

We therefore seek conditions on $M_\gamma$ such that all relevant singular values are nonvanishing and scale linearly with the total number of measurements, $\sigma_i \propto N$.

\subsection{Defining notation}
To begin with, we need to define the notation for our specific problem. The target states $\rho$ are pure or close-to-pure in a $d$-dimensional Hilbert space. The target state can be written in terms of an eigendecomposition (keeping the small eigenvalues for later convenience),
\begin{equation}
    \rho = \sum_{k=1}^d \lambda_k \ket{\lambda_k}\bra{\lambda_k},
    \label{eq:eigen-decomposition_appendix}
\end{equation}
where $\{\ket{\lambda_k}\}$ forms a basis in the $d$-dimensional Hilbert space. We still want to use full-rank estimators, which means that there are $d^2-1$ parameters that must be estimated, and $\nu = d^2-1$ in Eq.~\eqref{eq:app_mean_inf_singular_values}. We will purify this system by introducing an auxiliary Hilbert space of the same dimension, totaling $\mathcal{H}_d \otimes\mathcal{H}_d$. This purification allows us to represent the density matrices as state vectors. In addition to the purification, we will also express the purified complex state vector and the measurement operators in a space containing only real components, allowing us to write
\begin{equation}
    \begin{split}
        \rho_{d\times d}& \stackrel{
    \text{Purification}}{\rightarrow} \ \sum_{k=1}^d\sqrt{\lambda_k}\ket{k}_d\otimes\ket{\lambda_k}_d = \ket{\Psi}_{d^2} \\&\stackrel{
    \text{Real}}{\rightarrow}  v=\begin{pmatrix}
        \mathcal{R}(\ket{\Psi})\\
        \mathcal{I}(\ket{\Psi})
    \end{pmatrix},\\
    M_{\gamma, d\times d}& \stackrel{
    \text{Purification}}{\rightarrow} \mathbb{1}_{d\times d} 
    \otimes M_{\gamma, d\times d} = A_\gamma\\&\stackrel{
    \text{Real}}{\rightarrow} O_\gamma = \begin{pmatrix}
        \mathcal{R}(A_\gamma) & -\mathcal{I}(A_\gamma) \\
        \mathcal{I}(A_\gamma) & \mathcal{R}(A_\gamma)
    \end{pmatrix}.
    \end{split}
\end{equation}
Using this notation, we can rewrite the Born rule as 
\begin{equation}
    p_\gamma = \Tr(M_\gamma \rho) = v^TO_\gamma v,
\end{equation}
where written out explicitly,
\begin{equation}
\begin{split}
    \ket{\Psi} = \begin{pmatrix}
    \sqrt{\lambda_1}\ket{\lambda_1}\\
        \vdots\\
    \sqrt{\lambda_d}\ket{\lambda_d}
    \end{pmatrix} = \begin{pmatrix}
        \sqrt{\lambda_1}\bra{1}\ket{\lambda_1}\\
        \vdots\\ 
        \sqrt{\lambda_1}\bra{d}\ket{\lambda_1}\\
        \sqrt{\lambda_2}\bra{1}\ket{\lambda_2}\\
        \vdots\\
        \sqrt{\lambda_d}\bra{d}\ket{\lambda_d}
    \end{pmatrix} 
    \\
  \text{and}  \quad     A_\gamma
=
\begin{pmatrix}
M_{\gamma} & 0 & \cdots & 0 \\
0 & M_{\gamma}& \cdots & 0 \\
\vdots & \vdots & \ddots & \vdots \\
0 & 0 & \cdots & M_{\gamma}
\end{pmatrix}.
\end{split}
\end{equation}
Using this notation, we next compute the Fisher information matrix.

\subsection{Fisher information matrix for a multinomial distribution}
\label{app:Compute_fisher_info_multinomial}
Our measurement follows the multinomial distribution, and the likelihood functions can be written in the purified notation as
\begin{equation}
    \mathcal{L}(v;n_\gamma)\propto \Pi_\gamma (v^T O_\gamma v)^{n_\gamma},
\end{equation}
where to simplify the notation, we limit ourselves to one measurement setting, which leads us to omit the index $\alpha$ that was previously used in Sec.~\ref{sec:optimal_stratergy_mechanism}.
The score function is therefore (dropping $n_\gamma$ as an argument)
\begin{equation}
    s(v) = \ln(\mathcal{L}(v))= \sum_\gamma n_\gamma \ln(v^T O_\gamma v).
\end{equation}
We compute the Fisher information matrix using the second formulation  in Eq.~\eqref{eq:Fisher_info_matrix_formulation}, recalling that $p_\gamma = v^TO_\gamma v$,
\begin{equation}
    \begin{split}
        \frac{\ln(\mathcal{L}(v))}{\partial v_i}=& \sum_\gamma \frac{n_\gamma}{p_\gamma} \frac{\partial p_\gamma}{\partial v_i},\\ 
        \frac{\partial^2 \ln(\mathcal{L}(v))}{\partial v_i \partial v_j}=& - \sum_\gamma \frac{n_\gamma}{p_\gamma ^2} \frac{\partial p_\gamma}{\partial v_i} \frac{\partial p_\gamma}{\partial v_j} + \sum_\gamma \frac{n_\gamma}{p_\gamma} \frac{\partial^2 p_\gamma}{\partial v_i \partial v_j},
    \end{split}
\end{equation} 
where
\begin{equation}
        \frac{\partial p_\gamma}{\partial v_i}= 2 \sum_a\left(O_\gamma\right)_{ia} v_a.
\end{equation}
Combining everything, we get the overall result
\begin{equation}
[\mathcal{I}(v)]_{ij}= \left\langle\sum_\gamma \frac{4n_\gamma}{p_\gamma^2} (O_\gamma v)_i(v^T O_\gamma)_j - \sum_\gamma \frac{n_\gamma}{p_\gamma} \frac{\partial^2 p_\gamma}{\partial v_i \partial v_j }\right \rangle.
\end{equation}
We compute the expectation value noting that for a multinomial distribution we have $\langle n_\gamma \rangle = Np_\gamma$, which gives
\begin{equation}
    [\mathcal{I}(v)]_{ij}= N\left(\sum_\gamma \frac{4}{p_\gamma} (O_\gamma v)_i(v^T O_\gamma)_j - \sum_\gamma \frac{\partial^2 p_\gamma}{\partial v_i \partial v_j} \right).
\end{equation} 
The last term vanishes because the probabilities have to be conserved $\sum_\gamma p_\gamma = 1$, which means that any sum of derivatives must vanish $\sum_\gamma \partial p_\gamma/\partial v_i = 0$. This leads to the final version of the Fisher information matrix,
\begin{equation}
    \mathcal{I}(v)= N\sum_\gamma \frac{4}{p_\gamma} O_\gamma vv^T O_\gamma,
    \label{eq:real_fisher_info_matrix}
\end{equation} 
and in the complex form
\begin{equation}
    \mathcal{I}(\ket{\Psi})= N\sum_\gamma \frac{4}{p_\gamma} A_\gamma \ket{\Psi}\bra{\Psi} A_\gamma.
    \label{eq:Multinomial_Fisher_matrix}
\end{equation}

\subsection{Rank deficient estimator}
From the Fisher information matrix in Eq.~\eqref{eq:Multinomial_Fisher_matrix} and Eq.~\eqref{eq:real_fisher_info_matrix}, it is clear that the singular values $\sigma_i$ are generally proportional to $N$. However, this is not the case when the target state has fewer relevant degrees of freedom compared to the estimated state. When discussing the rank of the Fisher information matrix, we refer to the formulation in Eq.~\eqref{eq:real_fisher_info_matrix}. The full Fisher information matrix in Eq.~\eqref{eq:real_fisher_info_matrix} has rank $2d^2$. Half of these dimensions come from purification and are not physically relevant. The remaining dimensions are related to the physical degrees of freedom of the estimated state resulting in $d^2-1$ relevant dimensions, subtracting 1 for normalization. In the estimation problem outlined in Sec.~\ref{sec:defining_the_estimation_problem}, the target states are pure and therefore have $2(d-1)$ degrees of freedom. The rank deficiency of the estimation problem is the difference between the number of degrees of freedom in the estimator and the target state \cite{Struchalin2018},
\begin{equation}
\begin{split}
    \text{rank deficiency} &= d^2-1 -2(d-1) \\&\stackrel{d=2^n}{=} 4^n-1 -2(2^n-1).
    \label{eq:rank_deficiency}
    \end{split}
\end{equation}
 As an illustrative example, we will explicitly consider a two-qubit state. The target state can be written as
\begin{equation}
    \rho = \sum_{i=1}^4 \lambda_i\ket{\lambda_i}\bra{\lambda_i}
    \end{equation} 
with eigenvalues $\lambda_1 \approx 1$ and $\lambda_1\gg\lambda_2 \approx \lambda_3\approx \lambda_4 \approx 0$. This results in the terms of Eq.~\eqref{eq:Multinomial_Fisher_matrix} generically taking the form
\begin{equation}
    A_\gamma \ket{\Psi}\bra{\Psi} A_\gamma \approx \begin{pmatrix}
         M_\gamma\ket{\lambda_1}\bra{\lambda_1} M_\gamma   & 0 & 0 &0 \\
        0 & 0& 0 &0\\
        0 & 0& 0 &0\\
        0 & 0& 0 &0\\
    \end{pmatrix}.
    \label{eq:apsipsia_rank_defficient}
\end{equation} 
Therefore, due to the close-to-pure target state, the Fisher information matrix has rows and columns that are suppressed when all $p_\gamma \neq 0$. As a result, Eq.~\eqref{eq:app_mean_inf_singular_values} is not valid for the singular values that are small compared to the estimator uncertainty, and the reconstruction infidelity is dominated by contributions that scale suboptimally as $1/\sqrt{N}$. 

\subsection{Optimal adaptive measurement strategies}
Adaptive strategies recover optimal asymptotic scaling by amplifying the suppressed columns and rows of $\mathcal{I}(\ket{\Psi})$ by finding measurement operators $M_\gamma$ that satisfy $M_\gamma \ket{\lambda_k} =0$ for state vectors $\ket{\lambda_k}$ with nonvanishing eigenvalues. This effectively amplifies the suppressed rows and columns by a factor $1/p_\gamma$.  To see why this is the case, consider again the two-qubit example with a measurement operator of the form $M_\gamma = \tau_\gamma \ket{\psi_\gamma}\bra{\psi_\gamma}$, where $\tau_\gamma = \Tr(M_\gamma)$, and we define
\begin{equation}
    \ket{\psi_\gamma} = \alpha_{1,\gamma}\ket{\lambda_1} + \alpha_{2,\gamma}\ket{\lambda_2} + \alpha_{3,\gamma}\ket{\lambda_3} + \alpha_{4,\gamma}\ket{\lambda_4},
\end{equation}
where $\sum_i |\alpha_{i,\gamma}|^2  = 1$.
Using this form, we can write the outcome probabilities
\begin{equation}
    p_\gamma =\tau_\gamma \sum_i \lambda_i |\alpha_{i,\gamma}|^2 = \bra{\Psi} A_\gamma \ket{\Psi},
    \label{eq:gamma_probability}
\end{equation}
and the Fisher information matrix elements starting from
\begin{equation}
    A_\gamma \ket{\Psi} = \begin{pmatrix}
        \sqrt{\lambda_1}M_\gamma \ket{\lambda_1}\\
        \sqrt{\lambda_2}M_\gamma \ket{\lambda_2}\\
        \sqrt{\lambda_3}M_\gamma \ket{\lambda_3}\\
        \sqrt{\lambda_4}M_\gamma \ket{\lambda_4}
    \end{pmatrix} =\tau_\gamma \begin{pmatrix}
        \sqrt{\lambda_1}\alpha^*_{1,\gamma}\ket{\psi_\gamma}\\
        \sqrt{\lambda_2} \alpha^*_{2,\gamma} \ket{\psi_\gamma}\\
        \sqrt{\lambda_3} \alpha^*_{3,\gamma} \ket{\psi_\gamma}\\
        \sqrt{\lambda_4} \alpha^*_{4,\gamma} \ket{\psi_\gamma}\\
    \end{pmatrix},
        \label{eq:Apsi}
\end{equation}
which results in
\begin{equation}
   (A_\gamma \ket{\Psi})_i(\bra{\Psi} A_\gamma)_{j} = \sqrt{\lambda_i\lambda_j}\alpha_{i,\gamma}\alpha^*_{j,\gamma} \tau^2_\gamma \ket{\psi_\gamma}\bra{\psi_\gamma}.
    \label{eq:ApsipsiA}
\end{equation}
In Eq.~\eqref{eq:apsipsia_rank_defficient} only the upper left block remains nonvanishing when $\lambda_1\gg\lambda_2 \approx \lambda_3\approx \lambda_4 \coloneq \lambda$. 
If we select a measurement operator $M_\gamma$ where $\alpha_{1,\gamma} = 0$, we get
\begin{equation}
\begin{split}
     &\frac{1}{p_\gamma} A_\gamma \ket{\Psi}\bra{\Psi} A_\gamma=\\&  \tfrac{1}{|\alpha_{2,\gamma}|^2 + |\alpha_{3,\gamma}|^2 + |\alpha_{4,\gamma}|^2 } \begin{pmatrix}
     0&0&0&0\\
0 &  |\alpha_{2,\gamma}|^2& \alpha_{2,\gamma}\alpha^*_{3,\gamma} & \alpha_{2,\gamma}\alpha^*_{4,\gamma}\\
       0 &  \alpha^*_{2,\gamma}\alpha_{3,\gamma} &  |\alpha_{3,\gamma}|^2 & \alpha_{3,\gamma}\alpha^*_{4,\gamma} \\ 
        0&  \alpha^*_{2,\gamma}\alpha_{4,\gamma}   & \alpha^*_{3,\gamma}\alpha_{4,\gamma}   &|\alpha_{4,\gamma}|^2
    \end{pmatrix} \otimes M_\gamma,
\end{split}
\end{equation}
where the small target state eigenvalues $\lambda$ were canceled by a corresponding factor in $p_\gamma$.
This matrix introduces a new nonvanishing singular value to the Fisher information matrix $\mathcal{I}(v)$, since both $M_\gamma$ and the lower $3\times3$ matrix are outer products of state vectors and therefore have rank 1. For a two-qubit system, the rank deficiency (Eq.~\eqref{eq:rank_deficiency}) is \mbox{$4^n-1-2(2^n-1) \stackrel{n=2}{=}9$}, and the lower $3\times3$ block can form up to nine linearly independent matrices. One such set is given by
\begin{equation}
\begin{split}
\ket{\psi_1} &= \ket{\lambda_2}, \\
\ket{\psi_2} &= \ket{\lambda_3}, \\
\ket{\psi_3} &= \ket{\lambda_4}, \\
\ket{\psi_4} &= \frac{1}{\sqrt{2}}\left(\ket{\lambda_2} + \ket{\lambda_3}\right), \\
\ket{\psi_5} &= \frac{1}{\sqrt{2}}\left(\ket{\lambda_2} + i\,\ket{\lambda_3}\right), \\
\ket{\psi_6} &= \frac{1}{\sqrt{2}}\left(\ket{\lambda_2} + \ket{\lambda_4}\right), \\
\ket{\psi_7} &= \frac{1}{\sqrt{2}}\left(\ket{\lambda_2} + i\,\ket{\lambda_4}\right), \\
\ket{\psi_8} &= \frac{1}{\sqrt{2}}\left(\ket{\lambda_3} + \ket{\lambda_4}\right), \\
\ket{\psi_9} &= \frac{1}{\sqrt{2}}\left(\ket{\lambda_3} + i\,\ket{\lambda_4}\right).
\end{split}
\end{equation}
Because all of these measurement operators are orthogonal to $\ket{\lambda_1}$ and yield linearly independent matrices, an adaptive measurement strategy can recover optimal asymptotic infidelity scaling.

This mechanism extends naturally to the higher-dimensional cases. In general, the requirement for adaptive measurements to achieve optimal scaling is that the measurement operators must be orthogonal to all eigenvectors corresponding to nonvanishing eigenvalues of the target state. Specifically, the condition is
\begin{equation}
\begin{split}
M_\gamma=\tau_\gamma\ket{\psi_\gamma}\bra{\psi_\gamma} \text{ such that } \bra{\psi_\gamma}\ket{\lambda_k} = 0 \\\text{ for all } k \text{ with } \lambda_k\neq0,
\label{eq:measurement_condition_for_optimal_scaling}
\end{split}
\end{equation}
and the number of such measurement operators required is given by \cite{Struchalin2018}
\begin{equation}
N_\mathcal{M} =
4^n - 1 - \big(R2^{n+1} - R^2 - 1\big),
\end{equation}
for target states with $R$ nonvanishing eigenvalues.

\subsection{Optimal adaptive measurement strategies with noise}
The condition described in Eq.~\eqref{eq:measurement_condition_for_optimal_scaling} explains how using multiple measurement settings can recover optimal asymptotic infidelity scaling for rank deficient estimators. We can now investigate how readout noise influences the optimality of adaptive strategies. We will consider the depolarizing channel introduced in Eq.~\eqref{eq:depol_channel}, which is restated here for convenience,
\begin{equation}
    \mathcal{E}^p_\text{depol}(A) = p\frac{\mathbb{1}}{2^n} \Tr(A)  + (1-p) A.
    \label{eq:app_depol_channel}
\end{equation}
When this is applied to measurement operators of the form $M_\gamma = \tau_\gamma\ket{\psi_\gamma}\bra{\psi_\gamma}$, it gives
\begin{equation}
    \tilde M_\gamma  = \mathcal{E}_\text{depol}(M_\gamma)  = p\frac{\mathbb{1}}{2^n}\tau_\gamma  + (1-p)\tau_\gamma\ket{\psi_\gamma}\bra{\psi_\gamma}.    \label{eq:Depolarized_measurement_op}
\end{equation}
Computing the Fisher information matrix again using $\tilde M_\gamma$  in Eq.~\eqref{eq:Depolarized_measurement_op}, we have
\begin{equation}
A_\gamma \ket{\Psi} = \tau_\gamma
\begin{pmatrix}
\sqrt{\lambda_1}(\frac{p}{2^n} \ket{\lambda_1} + (1-p)\bra{\psi_\gamma}\ket{\lambda_1}\ket{\psi_\gamma})\\
\vdots \\
        \sqrt{\lambda_{2^n}}(\frac{p}{2^n} \ket{\lambda_{2^n}} + (1-p)\bra{\psi_\gamma}\ket{\lambda_{2^n}}\ket{\psi_\gamma})
    \end{pmatrix},
\end{equation}
and  
\begin{equation}
\begin{split}
  \left(\mathcal{I}(\ket{\Psi})\right)_{ij}  &= N\sum_\gamma \frac{4}{p_\gamma}\left(A_\gamma \ket{\Psi})_i(\bra{\Psi} A_\gamma \right)_{j} \\&=N\sum_\gamma \frac{4}{p_\gamma}\tau_\gamma^2 \sqrt{\lambda_i \lambda_j}a_{ij,\gamma}
    \label{eq:noisy_ApsipsiA}
\end{split}
\end{equation}
where $i,j \in \{1, \dots ,2^n\}$ and 
\begin{equation}
\begin{split}
        a_{ij,\gamma} =&  \left(\frac{p}{2^n} \ket{\lambda_i} + (1-p) \bra{\psi_\gamma} \ket{\lambda_i} \ket{\psi_\gamma}\right)\\ \times& \left( \frac{p}{2^n}\bra{\lambda_j} + (1-p) \bra{\lambda_j}\ket{\psi_\gamma} \bra{\psi_\gamma} \right).
\end{split}
\end{equation}
The outcome probabilities are lower bounded by
\begin{equation}
    p_\gamma = \tau_\gamma\frac{p}{2^n} + \tau_\gamma(1-p)\sum_{i=1}^{2^n}  \lambda_i |\bra{\psi_\gamma}\ket{\lambda_i}|^2 \geq \tau_\gamma\frac{p}{2^n},
\end{equation}
where we have used $ \lambda_i |\bra{\psi_\gamma}\ket{\lambda_i}|^2\geq 0$.
The finite lower bound on all $p_\gamma$ prevents the prefactor $1/p_\gamma$ in Eq.~\eqref{eq:noisy_ApsipsiA} from taking on sufficiently large values to recover suppressed rows and columns associated with small eigenvalues, regardless of the chosen measurement operator $M_\gamma$. Note that choosing a $\tau_\gamma\rightarrow 0$ will not help because $\tfrac{1}{p_\gamma} A_\gamma \ket{\Psi}\bra{\Psi} A_\gamma\propto \tau_\gamma$.

We emphasize that this bound is specific to full-rank estimators. The nonzero rank deficiency, Eq.~\eqref{eq:rank_deficiency}, forces the suppressed rows and columns of $\mathcal{I}(\ket{\Psi})$ to be recovered through the amplification factor $1/p_\gamma$ that the depolarizing floor $p_\gamma \geq \tau_\gamma p/2^n$ then caps. A rank-appropriate estimator ($R_e = R_s$) has no suppressed singular values to amplify, and the optimal $1/N$ scaling is retained even without adaptive measurements.

\section{Implementation details of Bayesian mean estimation}
\label{app:Implementation_BME}
The implementation of Bayesian mean estimation (BME) is primarily based on the structure suggested in Refs.~\cite{BlumeKohout2010, Struchalin2016}. The core idea of BME is to represent the likelihood function $\mathcal{L}(\rho)$ as an integrable posterior probability distribution $\pi_f(\rho)$, so that a mean estimate can be performed,
\begin{equation}
    \hat \rho_\text{BME} = \int d\rho\, \rho\, \pi_f(\rho).
\end{equation}
This is made computationally tractable by introducing a particle swarm, where the posterior distribution $\pi_f(\rho)$ is discretized and approximated by a set of states $\rho_i$ (``particles") and associated weights $w_i$, 
\begin{equation}
    \pi_f(\rho) \approx \sum_i \rho_i w_i. 
\end{equation}
This particle swarm representation forms the basis for implementing the Bayesian update cycle, which incorporates a data set $D = \{o_{f-1}, \dots ,o_1, o_0\}$ into the prior $\pi_0$, where $o_i$ is the index of the outcome of measurement $i$.

\subsection{The Bayesian update cycle}
The prior distribution $\pi_0(\rho)$ is initialized by drawing $n_\text{part}$ random particle states $\rho_i$ from the Hilbert-Schmidt distribution of mixed states \cite{yczkowski2011} and the weights equalized, $w_i = 1/n_\text{part}$. 
The data are integrated one by one through the Bayesian update rule, expressed in discretized form as
\begin{equation}
    w^{n+1}_i = \frac{w^n_i \Tr(\rho_i M_{o_{n+1}})}{\sum_j w^n_j \Tr(\rho_j M_{o_{n+1}}) },
\end{equation}
 where $M_{o_{n+1}}$ corresponds to the measurement operator for the $(n+1)$-th measurement outcome.

This update procedure is repeated until the majority of the weight is contained in only a few particles. When this happens, the approximation becomes unreliable, and the swarm must be resampled to better approximate the actual posterior distribution. To evaluate whether the distribution is too concentrated, we have used the following quantity,
\begin{equation}
    s = \frac{1}{\sum_i w_i^2},
\end{equation}
 where resampling occurs if $s<s_\text{thres} = \tau\times  n_\text{part}$, where $\tau$ is a threshold scaling. If this threshold is crossed, the distribution is resampled using the Metropolis-Hastings (MH) algorithm.

\subsection{Metropolis-Hastings resampling}
Resampling refines the particle swarm to better approximate the posterior distribution during updates. The key idea is to perturb existing particles so that they cluster more densely near regions of high posterior probability. The resampling follows the procedure from Appendix C of Ref.~\cite{Struchalin2016}, summarized here.

We randomly select a particle $\rho$ from the particle swarm, where the probability of selecting a particle $\rho_i$ is $w_i$. Then we uniformly perturb the particle $\rho \rightarrow \tilde \rho$, and accept the perturbation with probability $p$ using the following rule,
\begin{equation}
    p =
    \begin{cases}
        1 & \text{when }\mathcal{L}(\tilde \rho)\geq \mathcal{L}(\rho),\\
        \mathcal{L}(\tilde \rho) /\mathcal{L}(\rho), &\text{when } \mathcal{L}(\tilde \rho)<\mathcal{L}(\rho).
    \end{cases}
\end{equation}
If accepted, the particle state $\rho$ is updated to the perturbed state $\tilde \rho$. To achieve a uniform perturbation, we use a purification scheme in which the states are embedded in a larger Hilbert space $\rho_{d\cross d} \rightarrow \ket{\psi}_{d^2}$, where $d$ is the dimension of the original Hilbert space. We will drop the index in the following. In the purified space, the perturbed state vector $\ket{\tilde \psi}$ is obtained as follows,
\begin{equation}
    \ket{\tilde \psi} = a\ket{\psi} + b\frac{\ket{g}-\ket{\psi} \bra{\psi}\ket{g}}{||\ket{g}-\ket{\psi} \bra{\psi}\ket{g}||},
\end{equation}
where $a = 1-c^2/2$, $b = \sqrt{1-a^2}$, $c$ is drawn at random from the normal distribution $\mathcal{N}(0,\sigma)$, where $\sigma$ is proportional to the width of the current posterior distribution, and $\sigma =2k \sum_i (1-F(\rho_i, \bar \rho)) w_i$ where $\bar \rho = \sum_i w_i \rho_i$ and $k$ is a variance-strength modifier. $\ket{g}$ is a state vector with real and imaginary entries sampled from a normal distribution $\mathcal{N}(0,1)$. The purified perturbed state $\ket{\tilde \psi}$ is then traced back down to the original Hilbert space, where the step is accepted or rejected. The perturbation is then repeated $n_\text{MH}$ more times to the same state, before finally being inserted into a new particle swarm, where all weights are equalized $w_i=1/n_\text{part}$. This resampling procedure is repeated until $n_\text{part}$ new particles are inserted into the new particle swarm.

Explicit parameters used for the single- and two-qubit BME implementations are provided in Table~\ref{tab:BME_paramters} unless stated otherwise (the exception is Fig.~\ref{fig:qdt_main}, which uses \mbox{$n_{\text{part}} =2000$} for single-qubit estimation).
\begin{table}[]
    \centering
    \begin{tabular}{c|c|c}
        Parameter & One qubit & Two qubits  \\ \hline
         $n_\text{part}$ & 200 & 1000 \\
         $\tau$ & 0.1  & 0.05 \\
         $k$ & 0.15 & 0.4 \\
         $n_\text{MH}$  & 100 & 200\\

    \end{tabular}
    \caption{BME parameters that differ between single- and two-qubit cases.}
    \label{tab:BME_paramters}
\end{table}

\section{Supplementary numerical results}
\label{app:supplemental_restuls}

\subsection{Mixed target states}
\label{app:mixed_target_states}
\begin{figure}
 \centering
 \includegraphics[width=\linewidth]{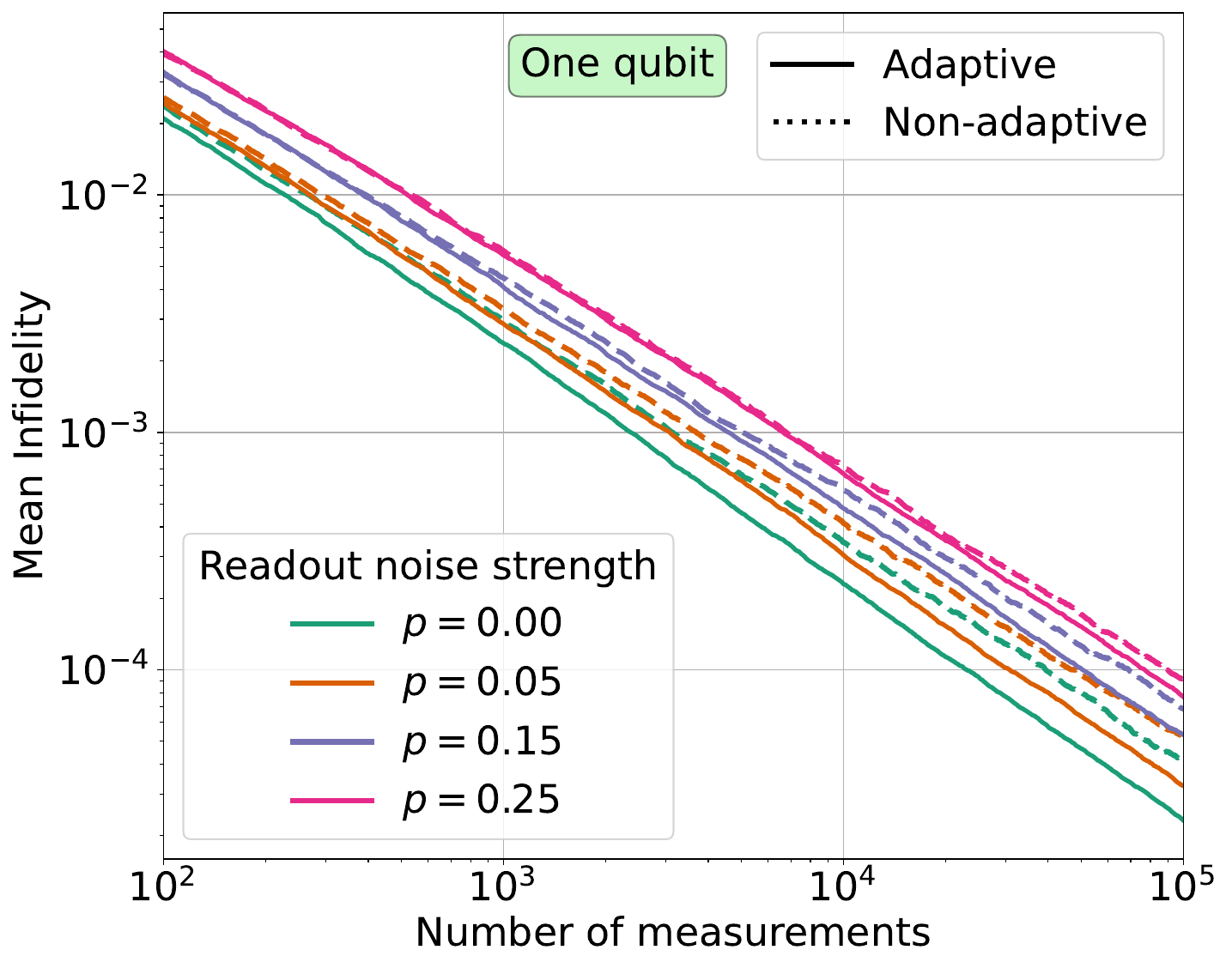}
 \caption{Mean reconstruction infidelity averaged over 1500 single-qubit Hilbert-Schmidt distributed states for varying depolarizing readout noise strength. All curves retain an approximate $1/N$ scaling over the entire range. Both axes are plotted on a logarithmic scale. }
 \label{fig:mixed_state}
\end{figure}

In Fig.~\ref{fig:mixed_state} we show the reconstruction infidelity for mixed target states with varying readout noise strength. The target states are drawn from the Hilbert-Schmidt distribution \cite{yczkowski2011}. Since mixed states do not suffer from the quadratic accuracy penalty of pure states, even nonadaptive methods achieve the ideal $1/N$ scaling. Performing adaptive measurements still improves the accuracy by an overall factor, but the improvement becomes smaller with stronger readout noise. At very high depolarizing strength adaptive and nonadaptive strategies become indistinguishable.

\subsection{Interquartile ranges}
\label{app:interquartile_range}

\begin{figure}
 \centering
 \includegraphics[width=\linewidth]{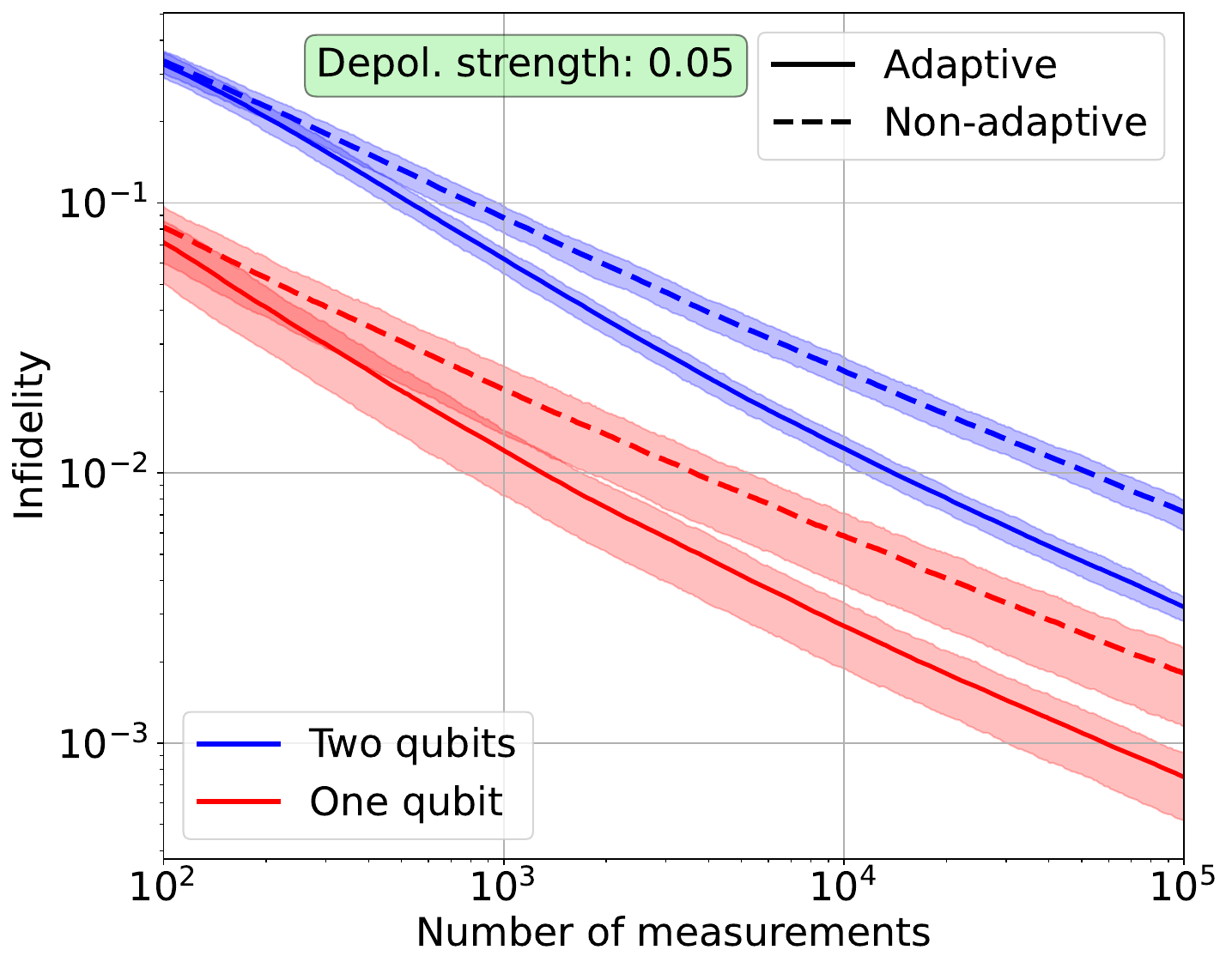}
 \caption{Interquartile range of single- and two-qubit infidelity curves with depolarizing readout strength $0.05$ from Fig.~\ref{fig:infidelity_comparison}. The shaded area denotes the range of the center $50\%$ of all curves at that measurement number. Both axes are plotted on a logarithmic scale.}
 \label{fig:interquartile_range}
\end{figure}

In Fig.~\ref{fig:interquartile_range} we show the interquartile ranges of the single- and two-qubit curves with depolarizing readout strength $p=0.05$ from Fig.~\ref{fig:infidelity_comparison}. The interquartile ranges are computed by sorting the infidelity at each measurement number and taking the center $50\%$ of the sorted points. We have consistent spread for both single- and two-qubit nonadaptive strategies, which is due to the random distribution of pure target states, and not to finite measurement statistics. When the target state is closer to the measurement basis, the state estimate performs better than when it is farther away, causing a natural spread \cite{Mahler2013}.

\subsection{Number of measurements to reach specified infidelity}
\label{app:shots_to_reach_infidelity}

\begin{figure}
 \centering
 \includegraphics[width=\linewidth]{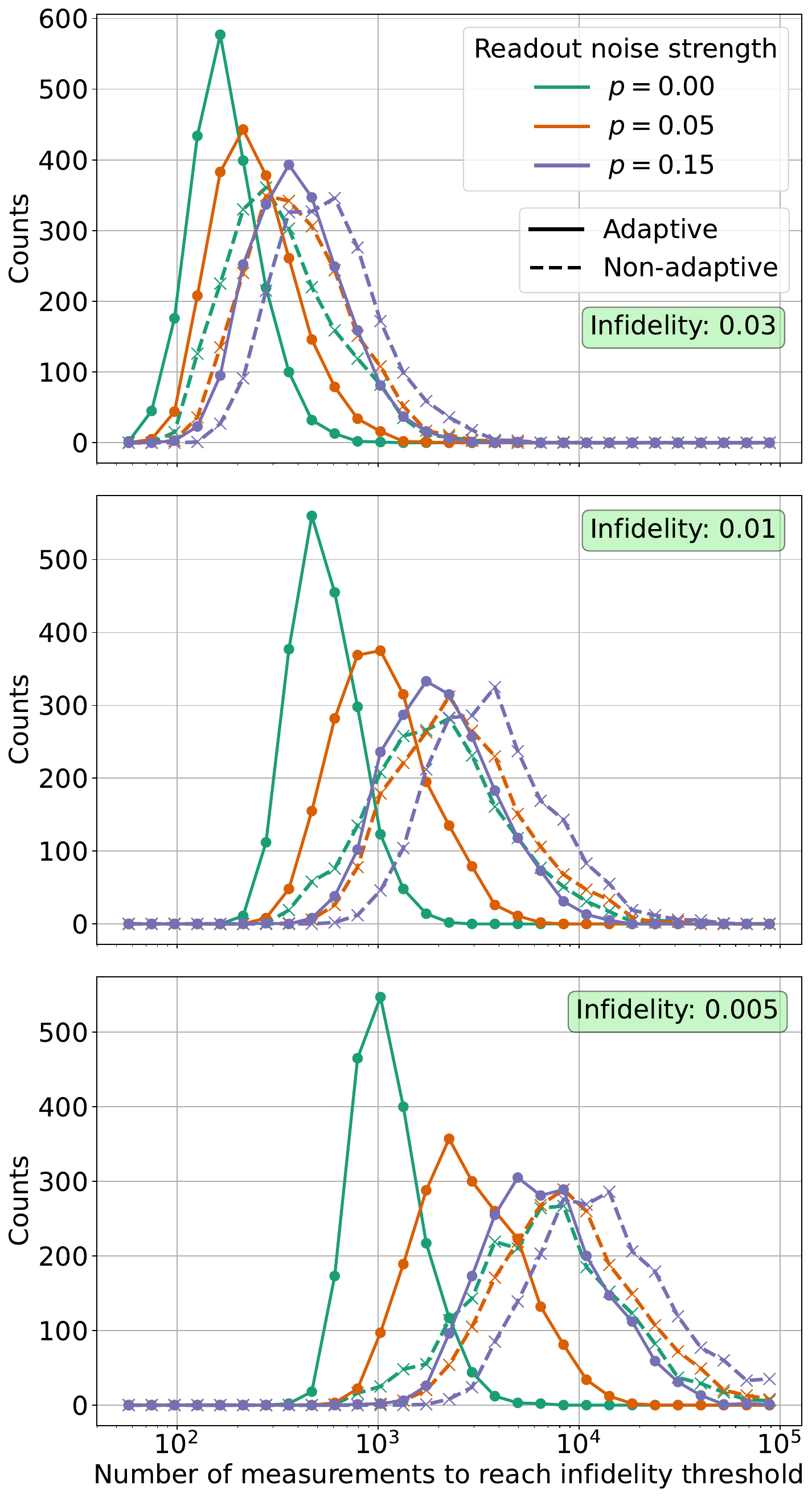}
 \caption{Binned distribution of the number of measurements required to reach a specified infidelity for the single-qubit curves in Fig.~\ref{fig:infidelity_comparison}. The $x$-axis is plotted on a logarithmic scale. }
 \label{fig:shots_to_reach_infidelity}
\end{figure}

In Fig.~\ref{fig:shots_to_reach_infidelity}, we show the binned distribution of the number of measurements required to reach a specified infidelity, corresponding to a vertical slice of Fig.~\ref{fig:infidelity_comparison}. The results are qualitatively similar to the binned distribution at a constant number of measurements in Fig.~\ref{fig:infidelity_at_shot_threshold}, which instead corresponds to a horizontal intersection of Fig.~\ref{fig:infidelity_comparison}. The noiseless adaptive distribution remains relatively peaked for all intersections, while the noisy adaptive cases broaden with increasing infidelity. The nonadaptive distributions are largely overlayered with slightly shifted peaks corresponding to the variance shift caused by readout-error mitigation. 

\section{Estimator uncertainty}
\label{app:Estimator_uncertainty}
\begin{figure}
 \centering
 \includegraphics[width=\linewidth]{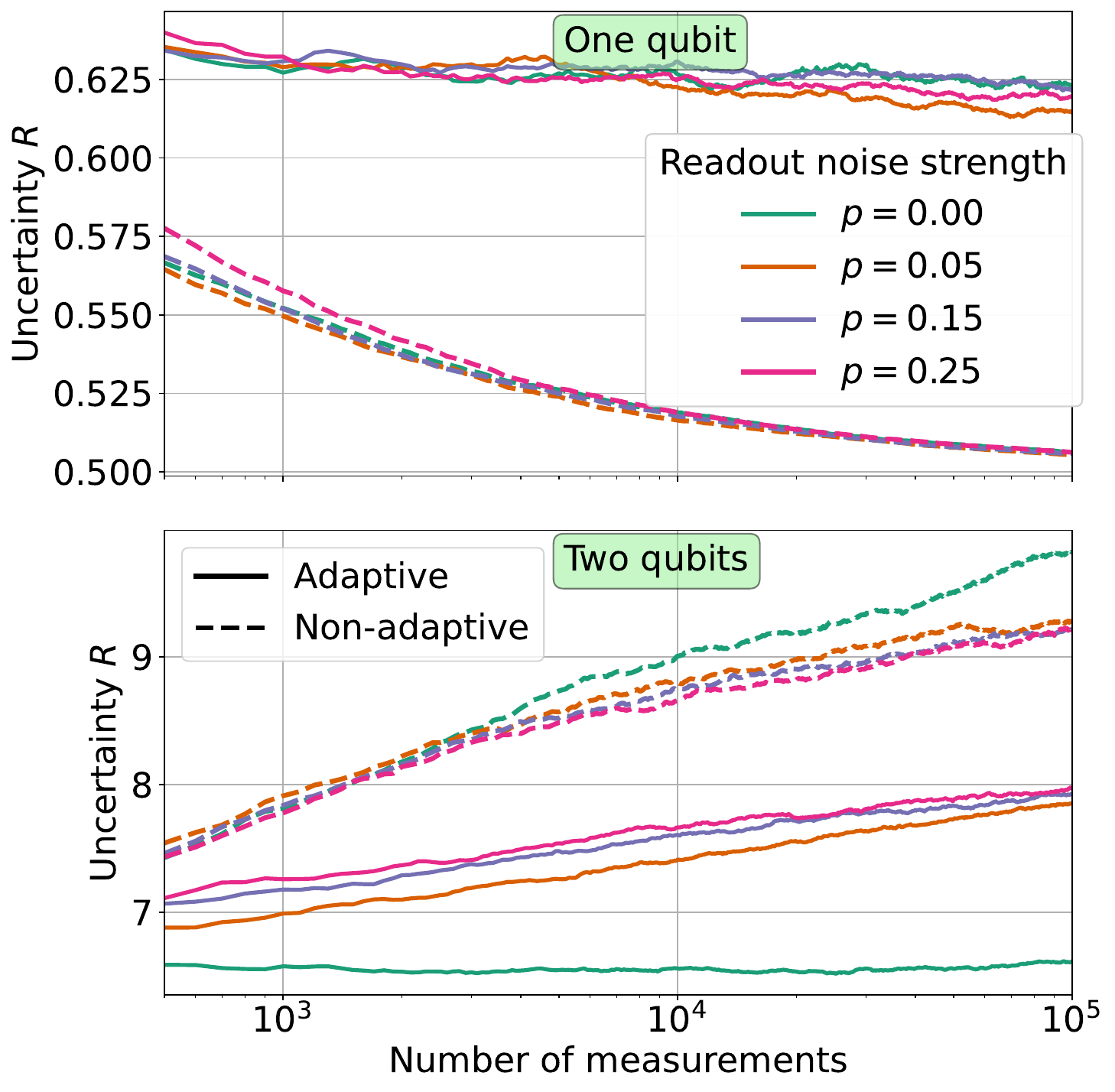}
 \caption{Single- and two-qubit estimator uncertainty. The uncertainty $R$ is computed according to Eq.~\eqref{eq:estimator_uncertainy}. The $x$-axis is plotted on a logarithmic scale. }
 \label{fig:uncertainty}
\end{figure}

To ensure that sufficiently many particles are used to estimate the Bayesian posterior distribution, we compute the relative estimator uncertainty,

\begin{equation}
    R = \frac{d_B(\hat \rho, \rho_T)}{\sum_i w_i d_B(\rho_i,\rho_T)},
    \label{eq:estimator_uncertainy}
\end{equation}
where $\rho_T$ is the target state, $\hat \rho$ is the current Bayesian mean estimate, and $\rho_i$ is the particle swarm with associated weights $w_i$, and 
\begin{equation}
    d_B(\rho, \sigma) = 2\left(1-\sqrt{F(\rho,\sigma)}\right)
\end{equation}
is the Bures distance. The relative uncertainty is not a complete measure of how reliable an estimator is, because it only gives a coarse sense of how far the distribution lies, relative to its spread, from the target state. The uncertainty should be about constant to ensure that the estimator converges correctly \cite{Struchalin2016}.  In Fig.~\ref{fig:uncertainty} single- and two-qubit uncertainties are plotted for the curves in Fig.~\ref{fig:infidelity_comparison}. In the single-qubit case the uncertainty is approximately constant for all noise levels. In the two-qubit case the noiseless adaptive uncertainty is approximately constant, while the others scale approximately logarithmically with the number of measurements. The growth of the uncertainty is not large, and is likely due to the shape of the likelihood function close to the estimation boundary.  Similar measurements can be found in Ref.~\cite{Struchalin2016}, where the authors check the uncertainty of estimators with Bures-random mixed target states, which we can reproduce with our estimator using the setting for the two-qubit BME in Fig.~\ref{fig:infidelity_comparison}, listed in Table~\ref{tab:BME_paramters}.

\begin{figure}
 \centering
 \includegraphics[width=\linewidth]{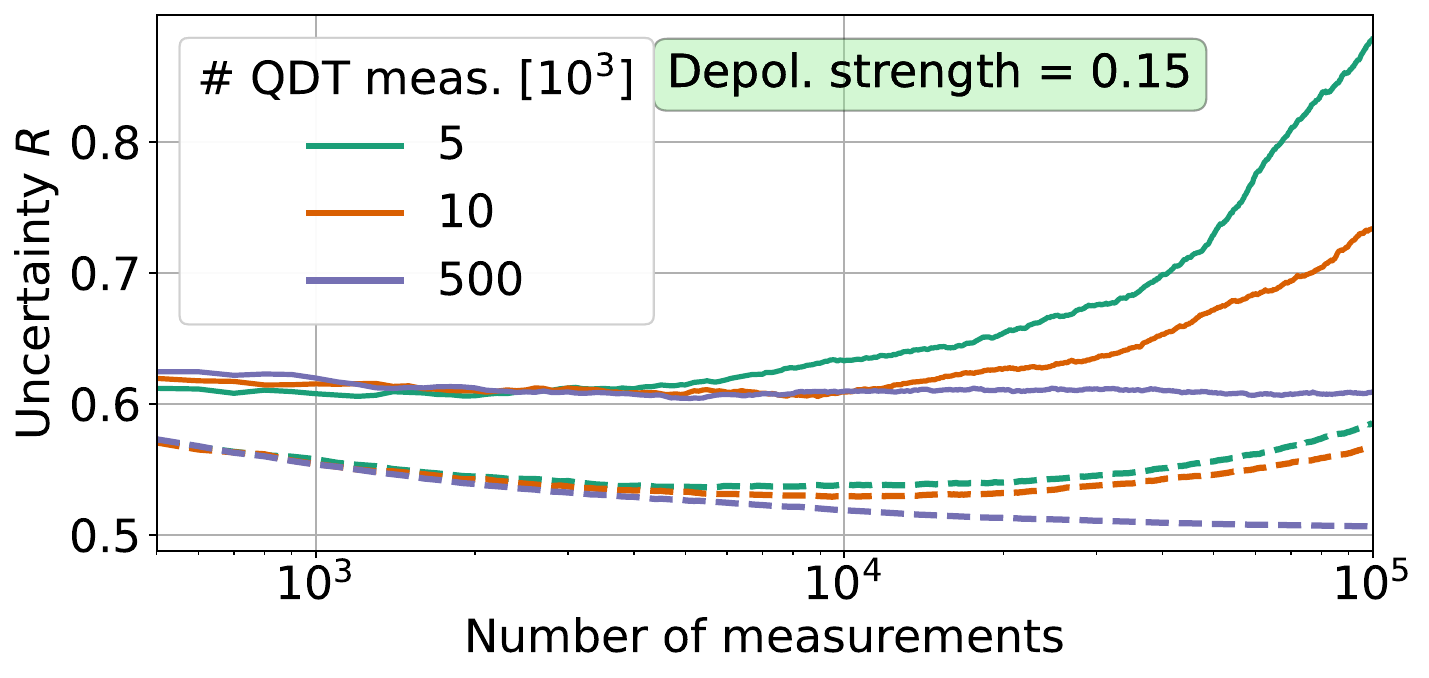}
 \caption{Estimator uncertainty from Fig.~\ref{fig:qdt_main} where a finite number of state copies were used for detector tomography. The curves using $5\times 10^3$ and $10\times 10^3$ grow superlogarithmically for the adaptive curves, and are therefore not robust estimators with these parameters. The uncertainty $R$ is computed with respect to Eq.~\eqref{eq:estimator_uncertainy}. The $x$-axis is plotted on a logarithmic scale. }
 \label{fig:qdt_uncertainty}
\end{figure}

In Fig.~\ref{fig:qdt_uncertainty} the uncertainties for the estimates in Fig.~\ref{fig:qdt_main} are plotted. We can see that the uncertainties remain about constant for all cases using more than  $10^4$ state copies for detector tomography. When fewer state copies are used, the uncertainty grows super-logarithmically, which means that the estimator does not converge properly, and the estimate should not be trusted in these cases. The estimator shown in Fig.~\ref{fig:qdt_uncertainty} uses a particle swarm with $n_\text{part}=2000$ states to have better convergence, in contrast to the standard choice of $n_\text{part}=200$ used in all other single-qubit estimators.

\section{Over- and underestimated depolarizing noise in detector tomography}
\label{app:qdt_appenidx}
\begin{figure}
 \centering
 \includegraphics[width=\linewidth]{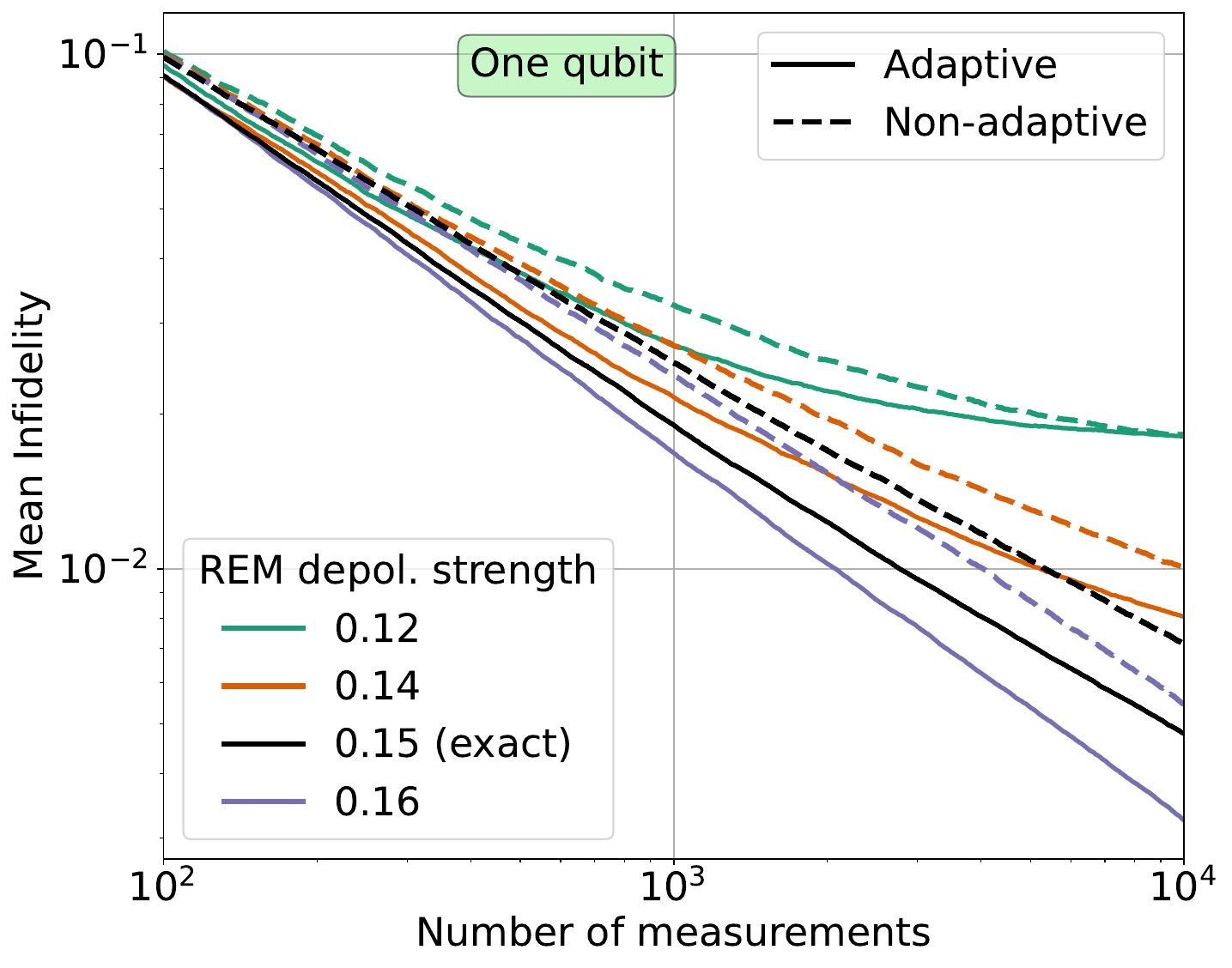}
 \caption{Mean infidelity curves with different depolarizing  noise strengths used for the measurement operators in readout-error-mitigated (REM) state tomography. The simulated target state is depolarized with strength $p=0.15$. The over- and underestimated cases are averaged over 1000 Haar-random states. The black curves use the exact depolarized measurement operators, and are equivalent to the curves used in Fig.~\ref{fig:infidelity_comparison}. Both axes are plotted on
a logarithmic scale.}
 \label{fig:qdt_appenidx}
\end{figure}

In Fig.~\ref{fig:qdt_appenidx}, we show readout-error-mitigated state tomography in which the measurement operators used for reconstruction include a constant depolarizing offset relative to those used to generate the data. Two distinct behaviors appear depending on whether the depolarizing strength is over- or underestimated. When the noise level is underestimated, the estimator converges to a state in the bulk of the estimator domain, introducing a constant bias for both adaptive and nonadaptive strategies. This is the general behavior expected from estimators with noise, such as the example in Fig.~\ref{fig:bias-variance}, where the unmitigated curves effectively underestimate the readout noise. 

In contrast, overestimating the depolarizing strength counterintuitively leads to better performance than using the exact error-mitigation parameters. This likely occurs because the target states are pure and lie on the boundary of the estimator domain, and when the readout noise is overestimated, it pushes the maximum of the likelihood function outside the physical domain, causing the estimator weight to accumulate rapidly on the boundary where the true state resides. This effect should disappear if the states are sufficiently mixed compared to the accuracy of the reconstructed measurement operators. The effect of over-estimating readout noise is not visible in Fig.~\ref{fig:qdt_main} because underestimated noise contributes significantly more to the infidelity averages.

Before discussing the behavior under overestimated noise in more detail, we emphasize that when measurement operators are poorly calibrated, a readout-error-mitigated estimator is not suitable for reliable state reconstruction. Our intent here is only to explain the qualitative features of the curves, not to advocate using such estimators in this regime.
When the readout noise is overestimated, the likelihood function is localized outside the estimator domain when overestimating the readout noise and the estimator does not converge well for these states. This is especially pronounced for the adaptive strategy, which saturates earlier than the nonadaptive strategy. This is likely because when the measurement setting keeps changing, the likelihood function is pushed outside the estimator domain in multiple directions, creating an inconsistent dataset for the estimator.

\bibliography{refs}

\end{document}